\documentclass{kapproc} 
\usepackage{graphicx}

\kluwerbib

\let\lcitebracket(
\let\rcitebracket)

\newcommand{\gray}{$\gamma$-ray}
\newcommand{\grays}{$\gamma$-rays}
\def\gtrsim{\mathrel{\hbox{\rlap{\hbox{\lower4pt\hbox{$\sim$}}}\hbox{$>$}}}}

\newcommand{\pubjournal}[5]{#1, {\bf #2}, #3, #4.}
\newcommand{\pubproc}[5]{#5, in {\it #1}, #2, #3, #4.}
\newcommand{\pubproca}[5]{#5, in {\it #1}, #2, #4.}
\newcommand{\aap}{Astron.\ Astrophys.}

\newcommand{\aph}{Astropart.\ Phys.}
\newcommand{\apj}{Astrophys.\ J.}
\newcommand{\apjl}{Astrophys.\ J.}
\newcommand{\apjs}{Astrophys.\ J.\ Supplement}
\newcommand{\mnras}{Mon.\ Not.\ Royal Astron.\ Soc.}
\newcommand{\nat}{Nature}
\newcommand{\na}{New Astron.}
\newcommand{\pasp}{Pub.\ Astron.\ Soc.\ Pacific}

\newcommand{\prd}{Phys.\ Rev.\ D}
\newcommand{\prl}{Phys.\ Rev.\ Lett.}

\newcommand{\icrc}{Int.\ Cosmic Ray Conf.}

\begin{document}

\articletitle{GLAST: Understanding the High Energy Gamma-Ray Sky}

\author{Julie E. McEnery$^*$}
\affil{NASA/Goddard Space Flight Center\\
Code 661, Greenbelt, MD 20771, USA }
\email{mcenery@milkyway.gsfc.nasa.gov}

\author{Igor V. Moskalenko\footnote{JCA/University of 
Maryland, Baltimore County, Baltimore, MD 21250, USA}}
\affil{NASA/Goddard Space Flight Center\\
Code 661, Greenbelt, MD 20771, USA }
\email{imos@milkyway.gsfc.nasa.gov}

\author{Jonathan F. Ormes}
\affil{NASA/Goddard Space Flight Center\\
Code 660, Greenbelt, MD 20771, USA }
\email{Jonathan.F.Ormes@nasa.gov}

\begin{abstract}

We discuss the ability of the GLAST Large Area Telescope (LAT) to
identify, resolve, and study the high energy \gray\ sky.  Compared
to previous instruments the telescope will have greatly improved
sensitivity and ability to localize \gray\ point sources.  The
ability to resolve the location and identity of EGRET unidentified
sources is described.  We summarize the current knowledge of the
high energy \gray\ sky and discuss the astrophysics of known and
some prospective classes of \gray\ emitters. Besides, we describe
the potential of GLAST to resolve old puzzles and to discover new 
classes of sources.

\end{abstract}


\section*{Introduction}

Our rudimentary understanding of the GeV \gray\ sky was greatly
advanced with the launch of Energetic Gamma-Ray Experiment Telescope
(EGRET) on the Compton Gamma-Ray Observatory (CGRO) in 1991. The
number of previously known GeV \gray\ sources increased from 1--2
dozen to the 271 listed in the 3rd EGRET Catalog. The \gray\ sky was
shown to be dominated by time-varying emitters. The science returns
from these observations exceeded prelaunch expectations. Among them
were the discoveries that for many blazars, intense \gray\ flares
are seen, that many blazars and some pulsars have peak luminosity at GeV 
energies, and that the spectrum of GRBs extends
to at least GeV energies. However, of this multitude of sources, only
101 have been definitively associated with known astrophysical
objects. Thus, most of the \gray\ sky, as we currently understand
it, consists of unidentified objects. This leaves intriguing puzzles
for Gamma-ray Large Aperture Space Telescope (GLAST), the next
generation GeV \gray\ instrument, to uncover.

One of the reasons that such a small fraction of the sources were
identified is because the size of the typical \gray\ error box from
EGRET was about 1 sq degree, an area that contains several candidate
sources preventing a straightforward identification. Most of the identified 
GeV sources have distinct temporal features that
allowed them to be associated with a known object. For example, the
observation of pulsation allowed unambiguous association between radio
pulsars and \gray\ sources, and the observation of correlated
radio, optical or X-ray emission with \gray\ flares allowed
the identification of several active galactic nuclei (AGN).
This is why most of the high
energy sources identified to date are either pulsars or AGN.

The large field of view and improved angular resolution of the LAT 
instrument on the GLAST mission will improve this situation
dramatically. The angular resolution and ``tails" on the point spread
function are improved compared to EGRET so we will be
able to refine maps of the \gray\ sky in crowded regions. The
higher effective area allows flares from AGN and pulsations from
pulsars and binary systems to be measured to correspondingly 
lower flux levels.

We begin with a brief description of the LAT as the next generation
instrument in the GeV energy range. We will then discuss in
Section~\ref{sec:current} the questions raised by the EGRET results
and the prospects for LAT studies of known  and potential 
classes of \gray\ sources and speculate on their numbers. 
Useful and up to date information can be found in GLAST Science 
Brochure\footnote{GLAST Science Brochure: 
http://glast.gsfc.nasa.gov/public/resources/brochures/}.

\section{Instrument description}
\subsection{Hardware}

\begin{figure}[tb!]
\vskip 5.3in  
\includegraphics{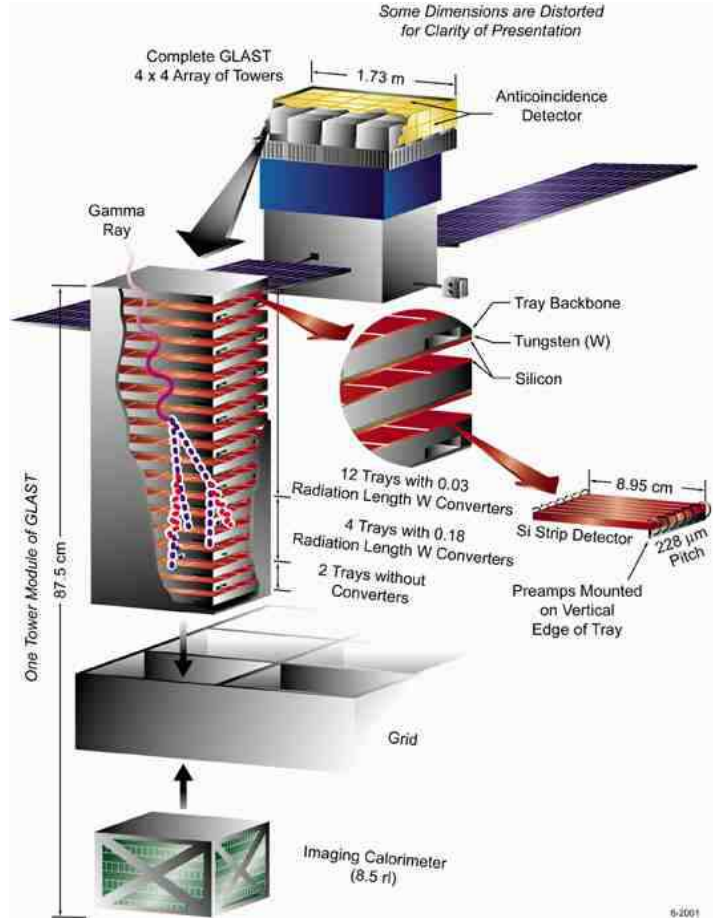}
\caption{The GLAST Large Area Telescope.}
\label{fig:glast}
\end{figure}

GLAST is a major space mission to explore the high energy \gray\
universe.  There are two instruments on board. The primary instrument
is the GLAST Large Area Telescope (LAT), which is sensitive at
energies from 20 MeV to 300 GeV. The secondary instrument is the GLAST
Burst Monitor (GBM)\footnote{GLAST Burst Monitor:
http://gammaray.msfc.nasa.gov/gbm}
to detect \gray\ bursts and provide broad--band
spectral coverage of this important phenomenon. The discussion of GBM
is given elsewhere. Like EGRET, the LAT will detect \grays\ through
conversion to electron-positron pairs and measurement of their
direction in a tracker and their energy in a calorimeter. It uses a
segmented plastic scintillator anti-coincidence system to provide
rejection of the intense background of charged cosmic rays. A
schematic of the LAT is shown in Fig.~\ref{fig:glast}.

Along with increased area with respect to EGRET, the design of each
element has been refined to improve the sensitivity and angular
resolution and to extend the energy range to higher energy.

In the tracker, incident photons convert to electron-positron pairs in
one of 16 layers of lead converter and are tracked by single-sided
solid-state silicon strip detectors (SSD) through successive planes
(the strips are centered every 228 microns).  Tracking the pairs
allows the reconstruction of the incident photon's arrival direction.
The technology is the same as that used for precision measurements at
the collision vertex in modern high-energy physics particle
investigations.  Multiple Coulomb scattering limits the angular
resolution in the 30 to 300 MeV band, but the geometry of the SSD
layers results in an important improvement over EGRET.  At higher
energies ($>$1 GeV) the precision of the EGRET spark chambers limited
the ability of that instrument to take advantage of the intrinsic
reduction in multiple scattering as energy increases. The very high
measurement resolution of the SSD being used in the LAT means that the
angular resolution is dominated by measurement uncertainty due to
multiple scattering only above energies of 100 GeV.  In determining
source positions, each photon can be weighted in proportion to its
energy, higher weight being given to higher energy particles whose
arrival direction is better known.  This results in an improvement in
angular resolution that can reduce the source location error boxes by
as much as a factor of 100 depending on the energy spectrum of photons
detected and the local \gray\ background.  Furthermore, LAT will
have 10--20 microsecond deadtime to enable for far better efficiency
for closely spaced (in time) \gray\ events during intense phases of
\gray\ bursts and solar flares.

A further subtlety in the LAT design can be found in the distribution 
of the thickness of the converter plates and SSD tracker planes. 
There are a total of 36 SSD planes, 18 $(x,y)$-pairs.  The first 12 pairs 
(the front section of the tracker) are covered by tungsten plates,
each of 0.03 radiation lengths (r.l.) thick, and the next 4 are 
covered by converters 0.18 r.l.\ thick, and the final 2 planes have 
no converter.  In addition, there are about 0.405 r.l.\ in silicon 
detector and structural materials.  The converter is thus 1.485 r.l.\
thick in total.  The photons that convert in the front section (0.36 
r.l.) give the highest angular resolution, while the back section 
assures that the overall photon detection efficiency is as high as 
practical.  The last two layers make sure we know the entry point of 
the particles into the calorimeter, where the tracking in the 
crystal layers, described below, continues. 

The calorimeter consists of a segmented arrangement of 1536 CsI (Tl)
crystals in 8 layers, giving both longitudinal and transverse
information about the energy deposition pattern. This has several
advantages over the shallower, monolithic calorimeter used in
EGRET. The segmentation allows for the shower development in the
calorimeter to be reconstructed resulting in a method of correction
for energy leaking out the bottom of the calorimeter. Combined
with the greater depth of the calorimeter this provides much the
required energy resolution at high energies up to several hundred
GeV. For normal incidence photons, shower maximum is contained up to
100 GeV. In addition, the segmentation allows the calorimeter to
provide direction information of the \gray\ photon which can be used if
the \gray\ photon did not convert in the tracker and provide important
pattern recognition inputs for the rejection of background.

The AntiCoincidence Detector (ACD) provides most of the rejection of
charged particle backgrounds. It consists of an array of 89 plastic
scintillator tiles each with charged particle detection efficiency of
0.9997 read out by wave-shifting fibers and photomultiplier tubes. The
segmentation prevents loss of effective area at high energy due to
self-veto caused by low energy photons coming from the cascade showers
in the calorimeter. This was an issue in EGRET that occured when
albedo from showers in their calorimeter fired the monolithic ACD
causing a 50\% reduction in effective area at 10 GeV. The ACD is being
designed so the LAT will have at least 80\% efficiency for 300 GeV
photons.

\begin{figure}[tb!]
\vskip 3.6in  
\includegraphics{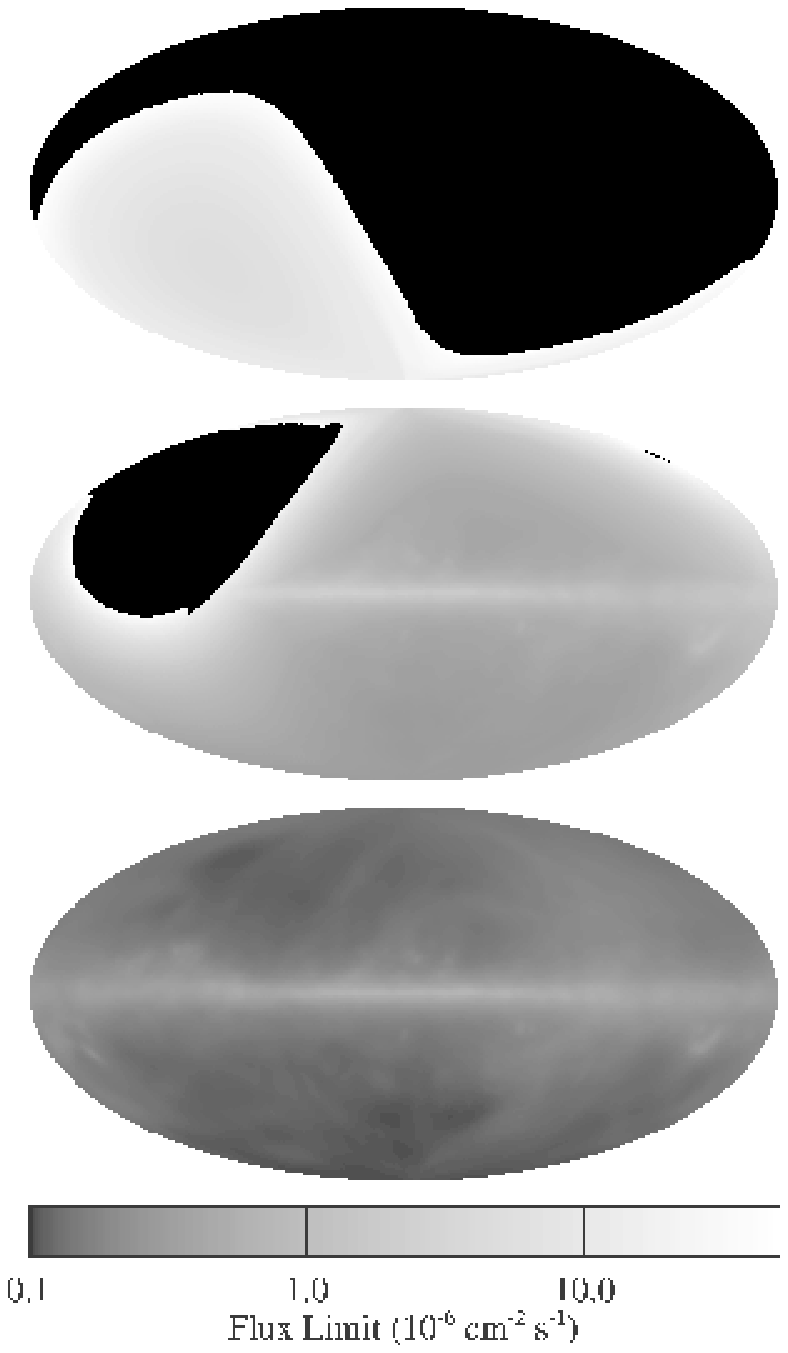}
\narrowcaption{Sensitivity and sky coverage of GLAST (top to bottom) 
for 100 s, one orbit, and one day observations.}
\label{fig:glastexp}
\end{figure}

The GLAST mission will be flown in low-Earth orbit and will operate in
both a ``rocking'' mode and a ``stare'' mode.  In rocking mode the
instrument moving 35 degrees north of zenith and then 35 degrees south
about the orbit plane on alternate orbits.  The instrument has a huge
field of view; $\sim$20\% of the sky at a time, and in rocking mode
covers $\sim$75\% of the sky every orbit. This is illustrated in 
Fig.~\ref{fig:glastexp} which shows the amount of sky coverage
and sensitivity in Galactic coordinates for 3 different integration
times while the instrument is in rocking mode. The top panel is for a
100 second observation. In rocking mode, the instrument does not move
significantly during this time, so the extent of the non-black region
indicates the size of the region of the sky that will be visible to
the LAT at any instant. The middle panel shows the sensitivity after a
one orbit (90 minute) observation.  After two orbits (one with GLAST
rocked north, and one rocked south) there is complete sky coverage.
For a one day observation (bottom panel) the exposure on the sky
becomes fairly uniform, the variations in point source sensitivity are
dominated by the distribution of the background due to the Galactic
diffuse emission. The sensitivity achieved after a single day
observation is similar to the point source sensitivity of EGRET for
the entire mission.

\subsection{Instrument capabilities}

\begin{figure}[tb!]
\vskip 1.9in
\includegraphics{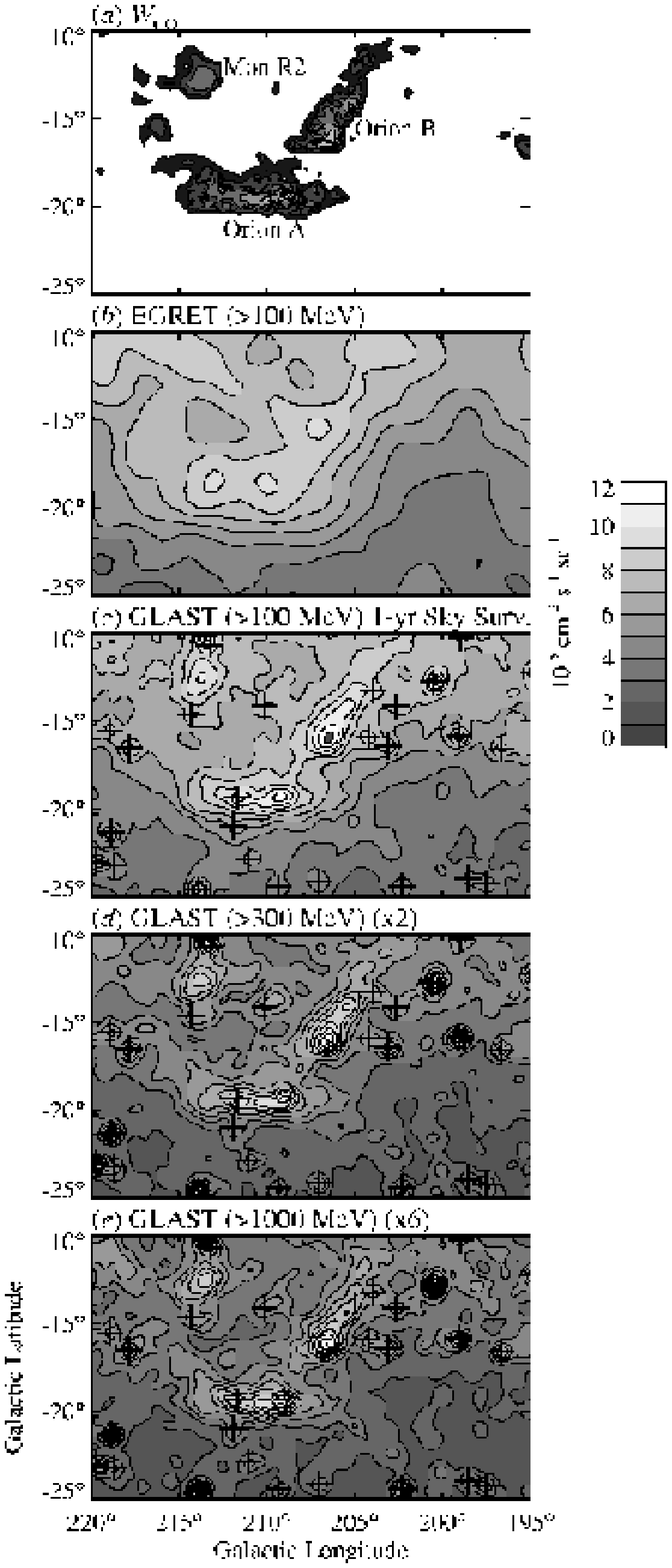}
\narrowcaption{(a) Integrated intensity of the 2.6-mm CO line in
Orion \cite{maddalena86}, showing the well-known Orion A and Orion B
molecular clounds and the Mon R2 cloud. (b) Intensity of \grays\
with energies $>$100 MeV observed by EGRET and analyzed
by Digel et al.~(1999). (c-e) Simulated intensity maps from the GLAST sky
survey for the energy ranges indicated. To smooth statistical
fluctuations, the EGRET data were convolved with a Gaussian of FWHM 
1.5$^\circ$, and the GLAST intensities with FWHM 0.75$^\circ$. Crosses
mark the positions of background sources with fluxes greater than $10^{-8}$
cm$^{-2}$ s$^{-1}$ ($>$100 MeV). The intensity scale refers to (b-e) with
scale factors as noted for (d-e) \cite{digel00}.}
\label{fig:orion}
\end{figure}

The capabilities of GLAST's LAT are summarized in
Table~\ref{tab:props}. 
The improvement in angular resolution will result in a much greater
ability to localize sources. The area of source location error boxes
will be reduced, compared to those of EGRET, by at least a factor of
three depending on the source spectrum. This means that GLAST will be
much more capable of resolving structure in the interstellar emission
and separating point sources from extended or diffuse sources of
emission.  This will be of particular interest in, for example,
locating and separating hot spots in supernova remnants 
from possible central point
sources or resolving the positions of giant molecular clouds.
Fig.~\ref{fig:orion} shows a simulation of the Orion region where
point sources and extended diffuse emission may be intermingled.

\begin{table}[t!]
\caption[Properties of the LAT compared to EGRET.]
{Properties of the LAT compared to EGRET.}
\begin{tabular*}{\textwidth}{@{\extracolsep{\fill}}lll}
\sphline
              &  EGRET  & LAT\cr
\sphline
Energy range &  20 MeV -- 30 GeV & 20 MeV -- 300 GeV \cr
Energy resolution & 10\%  & 9\% \cr
Effective area & 1500 cm$^2$ & 10000 cm$^2$ \cr
Angular resolution & 5.8$^\circ$ -- 0.3$^\circ$ & 3.4$^\circ$ -- 0.09$^\circ$ \cr
Field of view & 0.5 sr & 2.4 sr\cr
\sphline
\end{tabular*}
\label{tab:props}
\end{table}

\begin{figure}[b!]
\vskip 2.6in
\includegraphics{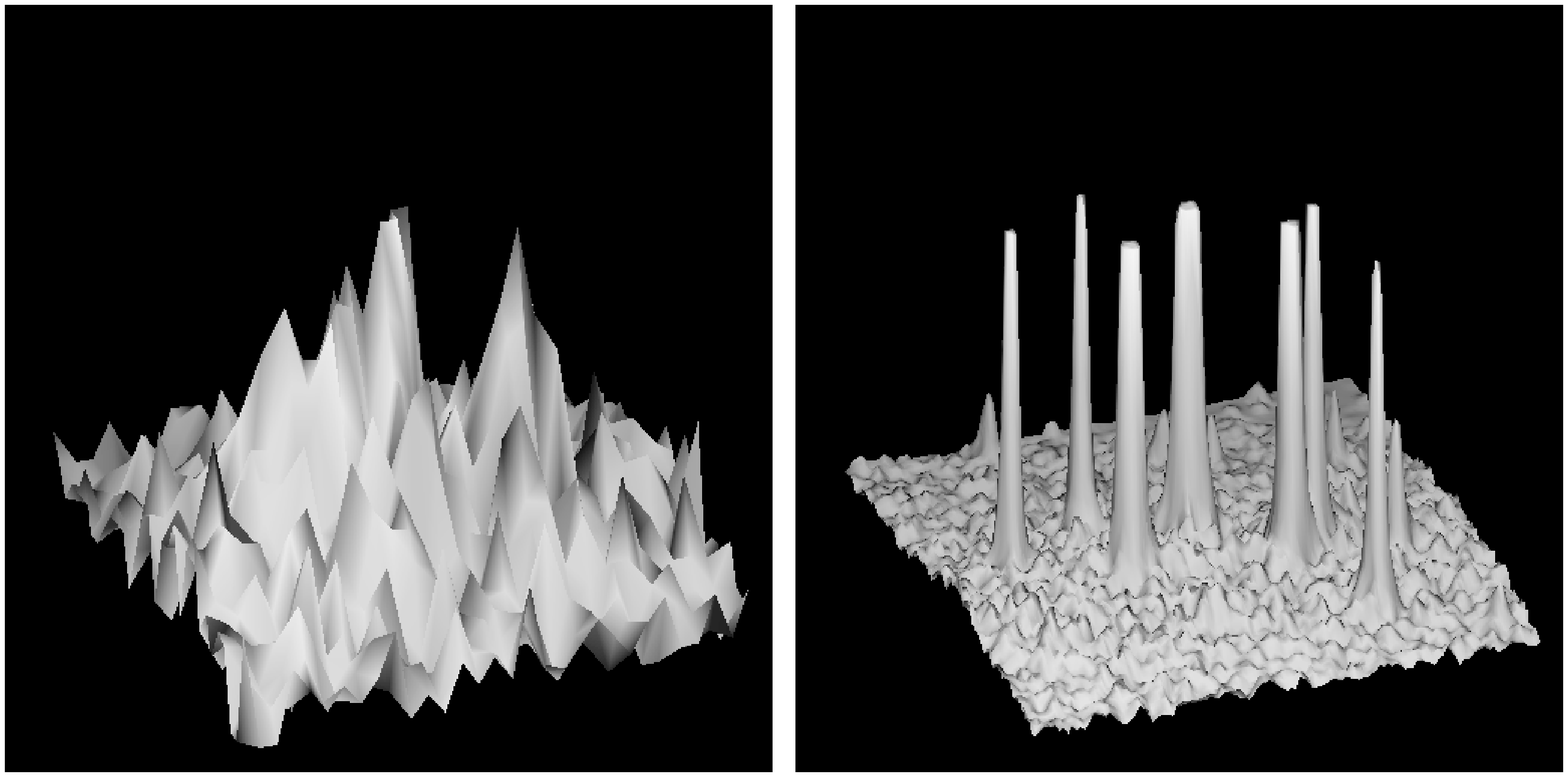}
\caption{Cygnus region simulations (S.~Digel, private communication).}
\label{fig:cygnus}
\end{figure}

GLAST will also fare far better in regions that were beset by problems of
source confusion (where the sensitivity becomes limited by background due to 
unresolved sources) in the EGRET all-sky survey.
Fig.~\ref{fig:cygnus} shows a comparison of the EGRET and simulated
LAT response for the Cygnus region. This illustrates the dramatic improvement
in source identification that will be possible with the superior source
localization abilities of the LAT.

The increase in effective area will be of enormous importance for
identification of sources via their temporal signatures. Flares from blazars
will be detected at much lower flux levels and on far shorter time
intervals. This greatly increases the likelihood of observing
correlated variability in several wavebands, allowing a firm
identification of the \gray\ source. The extra collection area will
also be very important for pulsar observations. The only Galactic
sources of high energy \grays\ that have been positively identified
are all young rotation-powered pulsars, so it is likely that many of
the unidentified sources are also pulsars. GLAST will be able to do
blind periodicity searches on all EGRET unidentified sources and thus
determine which are pulsars.  The LAT will have the potential to
detect \gray\ variability from new classes of putative \gray\
bright objects such as Galactic microquasars or dim, decaying signals
from \gray\ burst afterglows.

\begin{figure}[tb!]
\vskip 3.6in
\includegraphics{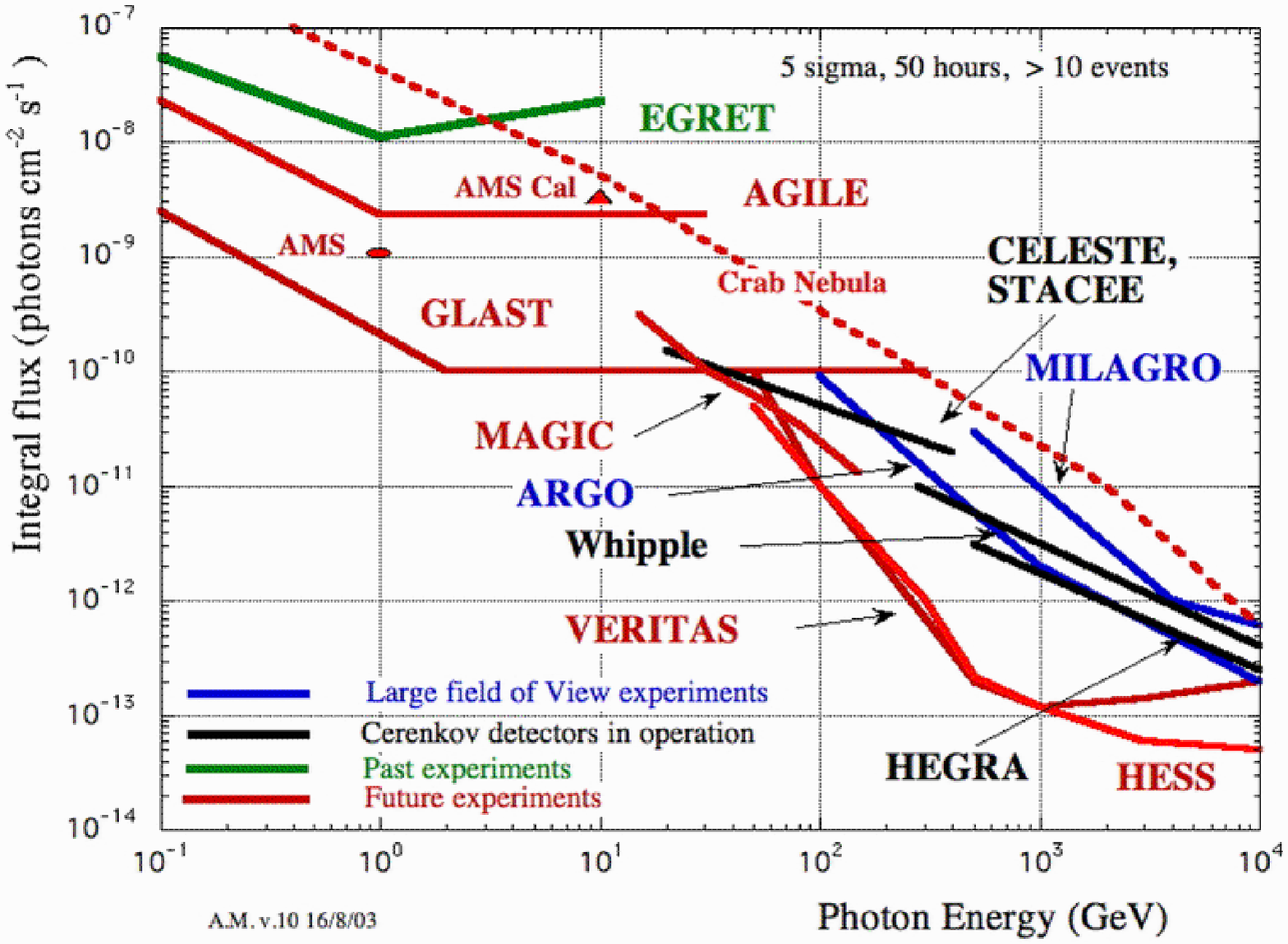}
\caption{Sensitivities of past, current and future high energy
\gray\ detectors \cite{Morselli}.}
\label{fig:sensall}
\end{figure}

The extended energy range opens important new discovery potential for
the LAT.  The EGRET effective area fell off rapidly above a few GeV;
the LAT has been carefully designed to maintain its area up to at
least 300 GeV.  This new high energy capability will provide
significant overlap with the next generation ground-based \gray\
telescopes, as illustrated in Fig.~\ref{fig:sensall}. It
shows the integral sensitivities of past, current and future high
energy \gray\ instruments.  This will provide complete spectral
coverage from MeV to TeV energies for the first time. It covers the
band over which we expect to see breaks in the spectra of
extragalactic objects due to the $\gamma\gamma$-absorption
on low energy photons in intergalactic space.  At the same time, the
calorimeter and its energy resolution capability allow searches for
\gray\ lines or other spectral features in the energy band where
the annihilation signatures of the
dark-matter candidate particles might be found.

This combination of GLAST's operational capability of scanning and
Earth avoidance and the LAT's increased field of view, effective area,
angular resolution and extended energy range in combination yield a
sensitivity about two orders of magnitude improvement in
sensitivity compared to EGRET.  This capability will be important to
do population and broadband spectral studies of such sources as
blazars, pulsars and gamma ray burst afterglows.

\section{Prospects: known and potential \gray\ sources}\label{sec:current}

There are only a few elementary processes capable of producing high
energy \grays\ (see Chapter 2 for more details), but uncovering their
relative efficiency provides us with important information about the
physical conditions and reactions in distant places that can not be
obtained by any other means.  The intensity of the \gray\ emission via
$\pi^0$-decay and bremsstrahlung depends on the target gas density,
while the intensity of inverse Compton (IC) scattering depends on the
density of photons.  Measurements of the \gray\ flux from an object
provides information such as the spectra and distributions of
accelerated particles, magnetic and radiation fields and gas density
and distribution. This information is vital for studies of most
astrophysical environments.

The nucleonic \grays\ are generated through the decay of
$\pi^0$-mesons produced in nucleus-nucleus interactions of accelerated
particles with gas.  Nucleonic \grays\ have a spectrum with maximum at
approximately $m_{\pi}c^2/2\approx 70$ MeV. For $E_\gamma\gg
m_{\pi}c^2/2$ the spectral index resembles the index of the ambient
energetic nucleons, $\alpha_p$.

The leptonic \grays\ can be produced via bremsstrahlung and IC
scattering of cosmic microwave background (CMB) photons, 
diffuse Galactic photons, and local
radiation fields.  Bremsstrahlung spectral index is approximately the
same as lepton's in the whole range, $\alpha_e$.  The spectrum of high
energy photons produced in IC scattering in the Thomson regime is
flatter than the spectrum of electrons,
$\alpha_\gamma=(\alpha_e+1)/2$.

The LAT is particularly well suited for observations of \grays\
produced by bremsstrahlung, IC scattering, and $\pi^0$-decay since its
energy range covers the regions where these processes play a major
role.  This orbital instrument together with a new generation of
ground-based telescopes will measure the \gray\ fluxes and spectra
with the high accuracy, required to provide insights into many
different types of astrophysical objects, and thus decipher
long-standing puzzles in cosmic-ray and \gray\ astrophysics.

This Section first describes the classes of known \gray\
sources and the major problems to be addressed by LAT,
and then speculates on the LAT potential to discover
the objects yet to be detected in \grays.

\subsection{Galactic diffuse \gray\ emission}
The diffuse continuum emission in the range 50 keV -- 50 GeV has been
systematically studied in the experiments OSSE, COMPTEL, EGRET on the
CGRO as well as in earlier experiments, SAS 2 and COS B. A review of
CGRO observations was presented by \cite{hunter97}. New models of the
diffuse emission are being developed by the LAT team. A new model will
be compared to the higher resolution data provided by the LAT and
important scientific understanding, described below, will result. 

\begin{figure}[tb!]
\vskip 3in
\includegraphics{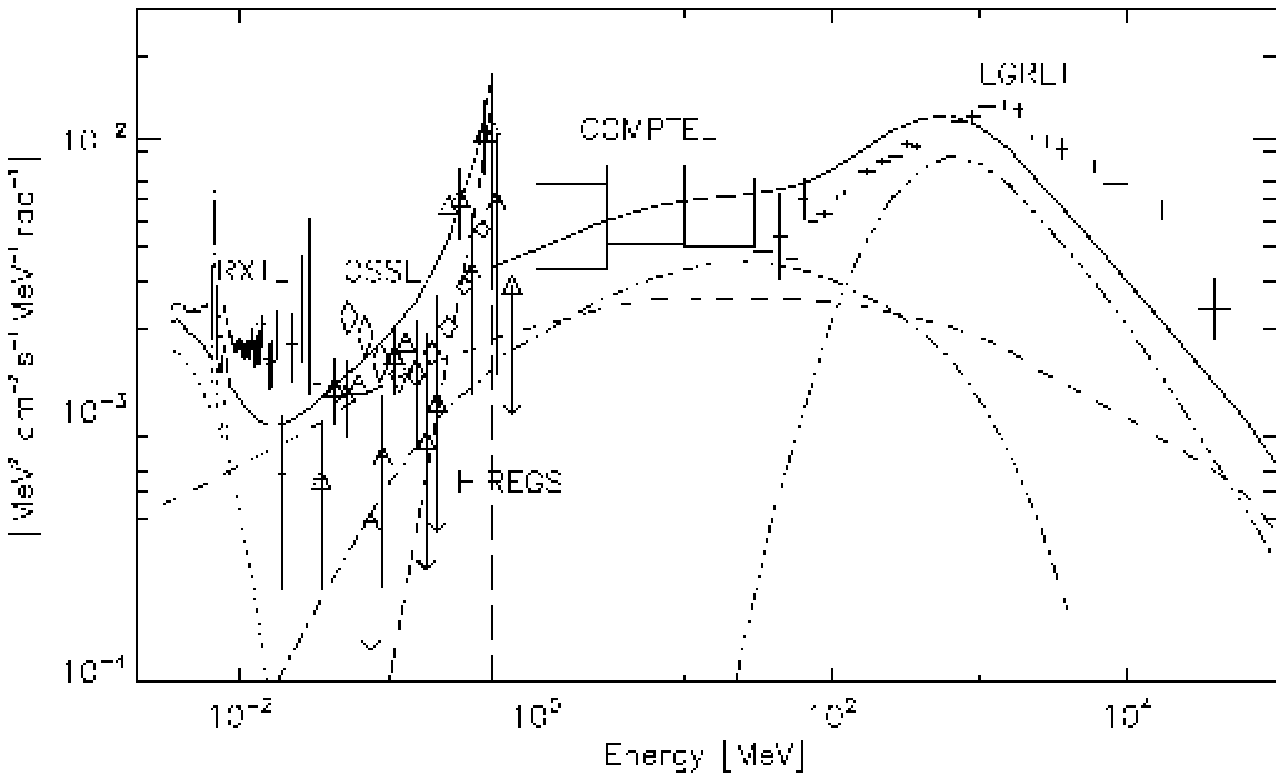}
\caption{Multiwavelength spectrum of Galactic diffuse \gray\ emission.
Hard X-ray/\gray\ Galactic diffuse emission as measured by
HIREGS (triangles), RXTE, OSSE/CGRO (diamonds), COMPTEL/CGRO, and
EGRET/CGRO. Also shown is the calculated flux (solid), and separate
components: bremsstrahlung (dash-dot), IC (short dash), and neutral
pions (triple dot-dash). Adapted from Boggs et al.~(2000).}
\label{fig:boggs}
\end{figure}

The diffuse \gray\ emission is the dominant feature of the \gray\ sky.  
The diffuse continuum \grays\ are produced in energetic interactions
of nucleons with gas via $\pi^0$ production, and by electrons via 
IC scattering and bremsstrahlung. In the plane of our Galaxy, the
emissions are most intense at latitudes below 5$^\circ$ and within
30$^\circ$ of the Galactic center and along spiral arms. Each of the
emission processes are dominant in a different energy range, and
therefore the \gray\ spectrum, can provide information about the
large-scale spectra of nucleonic and leptonic components of cosmic
rays (see Fig.~\ref{fig:boggs}). In turn, an improved understanding
of the role of cosmic rays is essential for the study of many topics
in Galactic and extragalactic \gray\ astronomy. It is worth noting
that an understanding of the spatial distribution and spectrum of the
diffuse emission is also important for studies of discrete sources.

A self-consistent model of Galactic diffuse \gray\ emission
should include cosmic-ray transport as the first step.
Knowing the number density of primary nuclei from satellite and
balloon observations, the production cross sections from the
laboratory experiments, and the gas distribution from astronomical
observations, one can calculate the production rate of secondary
nuclei. The observed abundance of radioactive isotopes determines then
the value of the diffusion coefficient, halo size and
other global parameters \cite{strong98}.

The modeling of cosmic-ray diffusion in the Galaxy includes the
solution of the transport equation with a given source distribution
and boundary conditions for all cosmic-ray species. The transport
equation describes diffusion, convection by the hypothetical Galactic
wind, energy losses, and possible distributed acceleration (energy
gain).  Electrons lose energy due to ionization, Coulomb scattering,
synchrotron emission, IC scattering, and bremsstrahlung.  The study of
transport of cosmic-ray nuclear component requires consideration of
nuclear spallation and ionization energy losses. The most
sophisticated analytical methods include disk-halo diffusion, the
dynamical halo Galactic wind, the turbulence and reacceleration (by
second order Fermi acceleration and by encounters with interstellar
shock waves).

The observation of diffuse \grays\ provide the most direct test of the
proton and electron spectra on the large scale. The EGRET observations
have confirmed the main features of the Galactic model derived from
locally observed cosmic rays; however, they also brought new
puzzles. The \grays\ revealed that the cosmic ray
source distribution required to match the cosmic ray gradient in the Galaxy 
should be distinctly flatter \cite{strong96} 
than the (poorly) known distribution of
supernova remnants \cite{case98}, the conventional sources of cosmic
rays. The spectrum of \grays\ calculated under the assumption that the
proton and electron spectra in the Galaxy resemble those measured
locally reveals an excess at $>1$ GeV in the EGRET spectrum
\cite{hunter97}. This may indicate that the cosmic ray proton and/or
electron spectra in the vicinity of the Sun are not representative of
the Galactic average.

Since the \gray\ flux in any direction is the line of sight integral,
attempts were made to explain the observed excess by a harder nucleon
spectrum in the distant regions \cite{mori97,gralewicz97}. However,
it seems that the harder nucleon spectrum is inconsistent with other
cosmic-ray measurements \cite{moskalenko98} 
such as secondary antiprotons and positrons, which are also
produced in $pp$-interactions.  While the large deviations in the proton
spectrum are less probable, the electron spectrum may fluctuate from
place to place. The rate of energy loss for electrons increases with
energy, it is thus natural to assume that the electron spectra are
harder near the sources producing more high-energy \grays\
\cite{porter97,pohl98,strong00}. IC scattering of
Galactic plane and CMB photons off electrons
provide a major contribution to the Galactic diffuse emission from
mid- and high-latitudes. The effect of anisotropic scattering in the
halo \cite{moskalenko00} increases the contribution of Galactic
\grays\ and reduces the extragalactic component.

New measurements of the spectrum of diffuse continuum Galactic 
\grays\ by GLAST will address several long-standing problems. The large
collection area and efficient operating mode will permit spectra to be
derived for much smaller area bins than was possible with EGRET. This
will allow for much better measurements of the latitude and longitude
distributions of the diffuse emission allowing better constraints to
be placed on its origin. The extension of the energy reach of the LAT
will allow the excess above 1 GeV found by EGRET to be
confirmed and characterized. This will greatly improve our
understanding of the character of cosmic-ray diffusion and
acceleration in our Galaxy and galaxies nearby. This in turn will
allow the determination of a better background model for both point
source and extended source studies.

More detailed discussion of Galactic diffuse emission may be found 
in Chapter 11. 

\subsection{Pulsars and plerions}

Pulsars make up the second most numerous class of identified sources
in the EGRET catalog which includes six confirmed and three candidate
\gray\ pulsars (see Chapter 7 for more details). Many of the
unidentified sources may be pulsars, however in many cases EGRET was
not able to collect enough photons per source to perform independent
period searches to detect these.

The lightcurves of all the EGRET pulsars show a double peak. Aside
from the Crab the shapes of the radio and \gray\ profiles are often
quite different with the peaks falling at different pulse phases.
This implies that low- and high-energy photons are most probably
emitted from different regions and thus that their origin is
different. Different mechanisms may be even responsible for emission
of low- and high-energy \grays\ \cite{mclaughlin00}, with the
efficiency of converting the spin-down energy to high-energy \grays\
increasing with pulsar age. In the case of Geminga pulsar, one of the
brightest \gray\ sources, most of its energy is emitted in GeV \grays\
\cite{jackson02} while its radio emission has not yet been detected.

The spectra of these objects are very hard; pulsed emission above 5
GeV was seen by EGRET for all six confirmed \gray\ pulsars. However,
spectral breaks are seen in most of these objects. PSR 1706--44
exhibits a break from a power-law spectrum with differential index
of --1.27 to a power-law spectrum with index --2.25 above 1 GeV. Vela,
Geminga, and the Crab all show evidence for a spectral
break at around one GeV. 
Stringent upper limits on PSR 1951+32 and PSR 1055--52 at a few
hundred GeV imply a spectral break for these objects \cite{srinivasan97}. 
However, there
were insufficient detected photons to allow a determination of the
shape of these spectral breaks. Searches for pulsed emission above a
few hundred GeV by ground-based \gray\ detectors have so far only
resulted in upper limits.

Two main types of models, polar cap and outer gap, have been proposed
to explain the pulsar \gray\ emission (see Chapter 3 for more
details).  These have been succesful in explaining some features of
\gray\ emission, but there is no model explaining all observations.
The \emph{polar cap} model \cite{daugherty96} explains the observed \grays\
in terms of curvative radiation or IC scattering of charged
particles accelerated in rotation-induced electric fields near the
poles of the pulsar.  The superstrong magnetic field in the pulsar
magnetosphere and the dense low-energy photon environment make it
highly probable that high-energy \grays\ are converted into
$e^\pm$-pairs.  To escape, \gray\ photons should be directed outwards
along the field lines. Another model, the \emph{outer gap} model
\cite{zhang97}, considers curvature \gray\ production by
$e^\pm$-pairs in the regions close to the light cylinder.  The pairs
are created in $\gamma\gamma$-interactions of high-energy photons with
thermal X-rays from the pulsar surface.

\begin{figure}[tb!]
\vskip 3.2in
\includegraphics{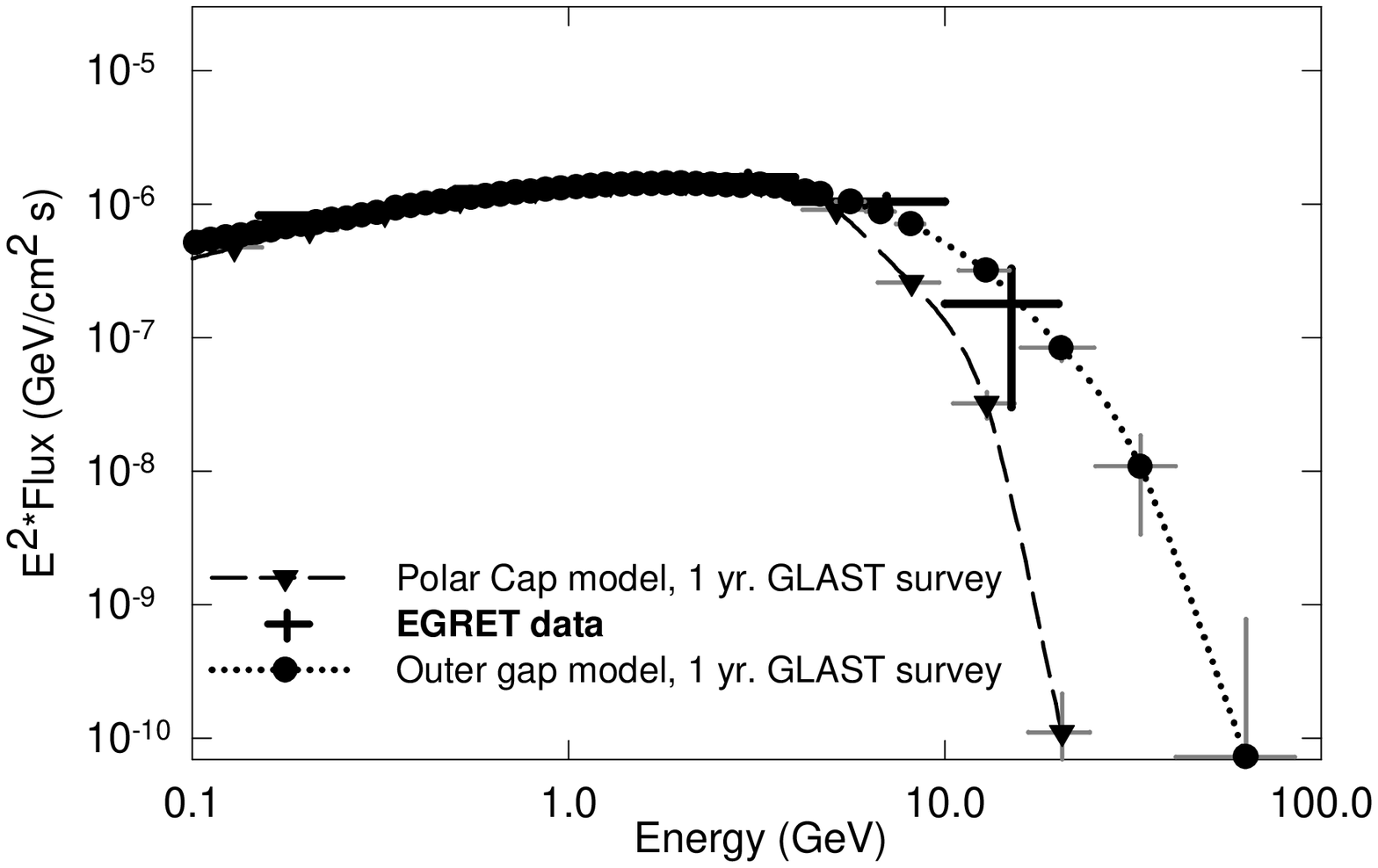}
\caption{High energy spectrum of the Vela pulsar. Heavy error bars: EGRET
data. Dotted line: outer gap model. Dashed line: polar cap model. Error
bars shown on the models are those expected from the GLAST mission in a 
one-year sky survey \cite[and references therein]{thompson01}.}
\label{fig:pulsar2}
\end{figure}

The high energy spectra of these two models are quite different, both
predict a spectral break at high energies, but the phase resolved
spectrum of the polar cap model falls off much more rapidly at high
energies (super-exponential) compared to the outer gap model
(exponential cutoff). There is also the possibility of a second, higher
energy, pulsed component due to IC scattering expected in
some outer gap models. Fig.~\ref{fig:pulsar2} illustrates how the LAT
high-energy response and spectral resolution will enable these two models 
to be easily distinuished for bright pulsars.

\begin{figure}[tb!]
\vskip 4in
\includegraphics{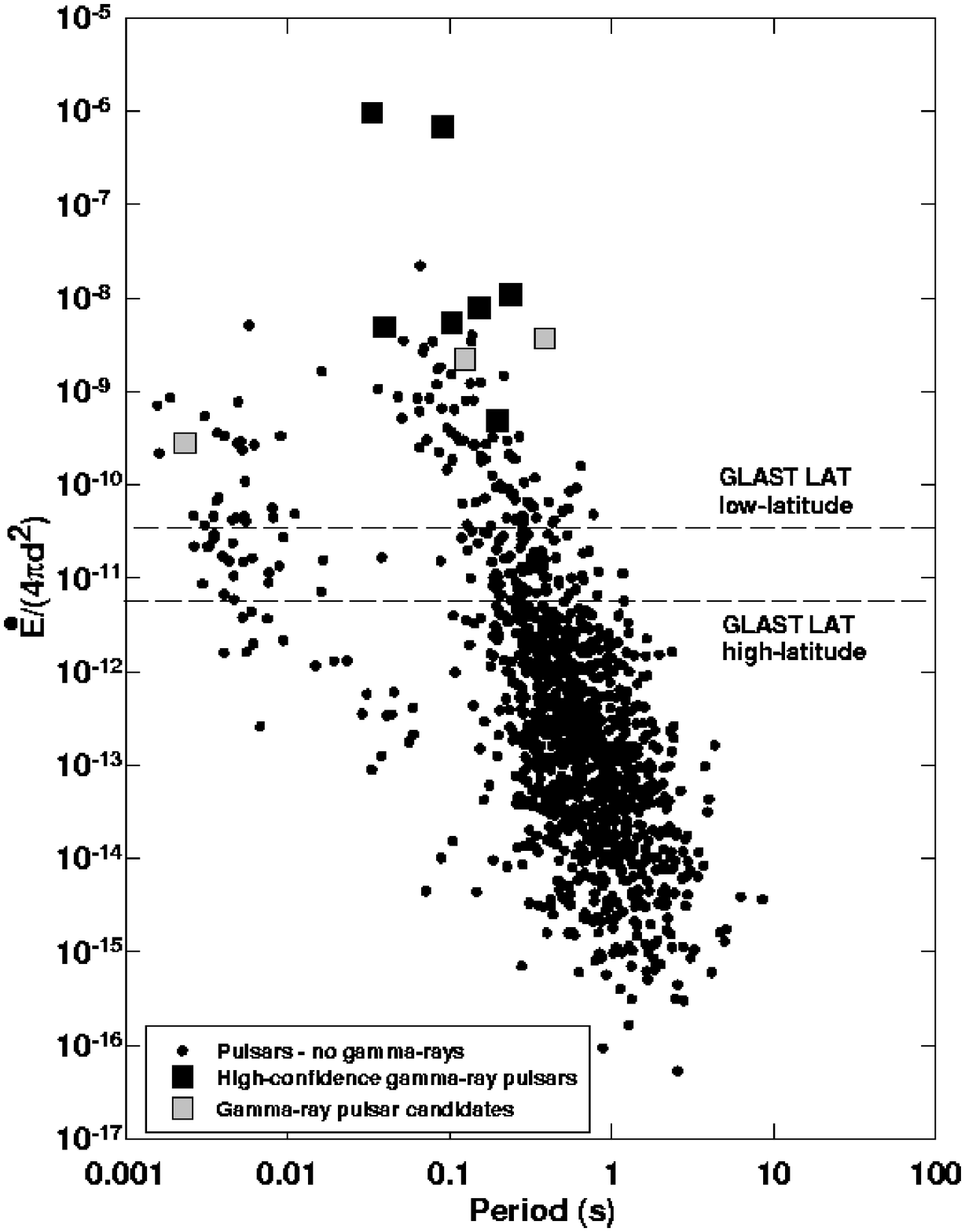}
\caption{Gamma-ray pulsar observability as measured by the spin-down energy 
seen at Earth (Thompson, Chapter 7).}
\label{fig:pulsar3}
\end{figure}

The two types of models also predict different ratios of radio-loud and
radio-quiet \gray\ pulsars. Polar cap models predict
a much higher ratio of radio-loud to radio-quiet \gray\ pulsars,
because in these models the high-energy and radio emission both
originate in the same magnetic polar region. Thus a measurement of
the ratio of radio-loud to radio-quiet \gray\ pulsars provides an
important discrimination between emission models. GLAST will detect
many more pulsars and thus will measure this ratio. The number of
pulsars that GLAST will see depends on the emission mechanism and on
the distribution of these sources on the sky. An empirical estimate
made by extrapolating a $\log N-\log S$ curve of the known pulsars suggests
that GLAST might expect to detect between 30 and 100 \gray\ pulsars.
Fig.~\ref{fig:pulsar3} shows one of the classic measures of pulsar
observability, the spin-down energy seen at Earth. Six of the seven
pulsars with the highest value of this parameter are \gray\
pulsars. The GLAST sensitivity will push the lower limit down
substantially.

Understanding the physics of pulsar \gray\ emission may be important
to determine the nature of the unidentified EGRET sources. 73 out of
170 unidentified sources are located in or near the Galactic plane,
and are likely to be pulsars or yet unknown source population.  Some
of them could be possibly associated with the Gould Belt, a massive
star Galactic structure surrounding the sun. A new population of
low-flux sources at an approximate distance 100--300 pc
\cite{gehrels00} may be misaligned young pulsars \cite{harding01}.

Observations with the LAT will improve our understanding of pulsars
in several ways. The high-energy response and energy resolution will
allow a determination of the shape of the high energy cutoffs in the 
bright pulsars. The increased collection area and thus improved photon
statistics will allow a search for periodicities on timescales of 
millseconds to seconds in sources as faint as $\sim$5$\times$10$^{-8}$
without prior knowledge from radio data, allowing for populations studies
with these sources.


Studying the pulsar driven nebulae is another way to understand the
physics of particle acceleration by pulsars. The Crab Nebula has been
detected at TeV energies by several groups \cite{konopelko98}.
There has also been reports of TeV emission from PSR 1706--44
\cite{kifune95}, and Vela \cite{yoshikoshi97} by the CANGAROO
collaboration.

\begin{figure}[tb!]
\vskip 2.85in
\includegraphics{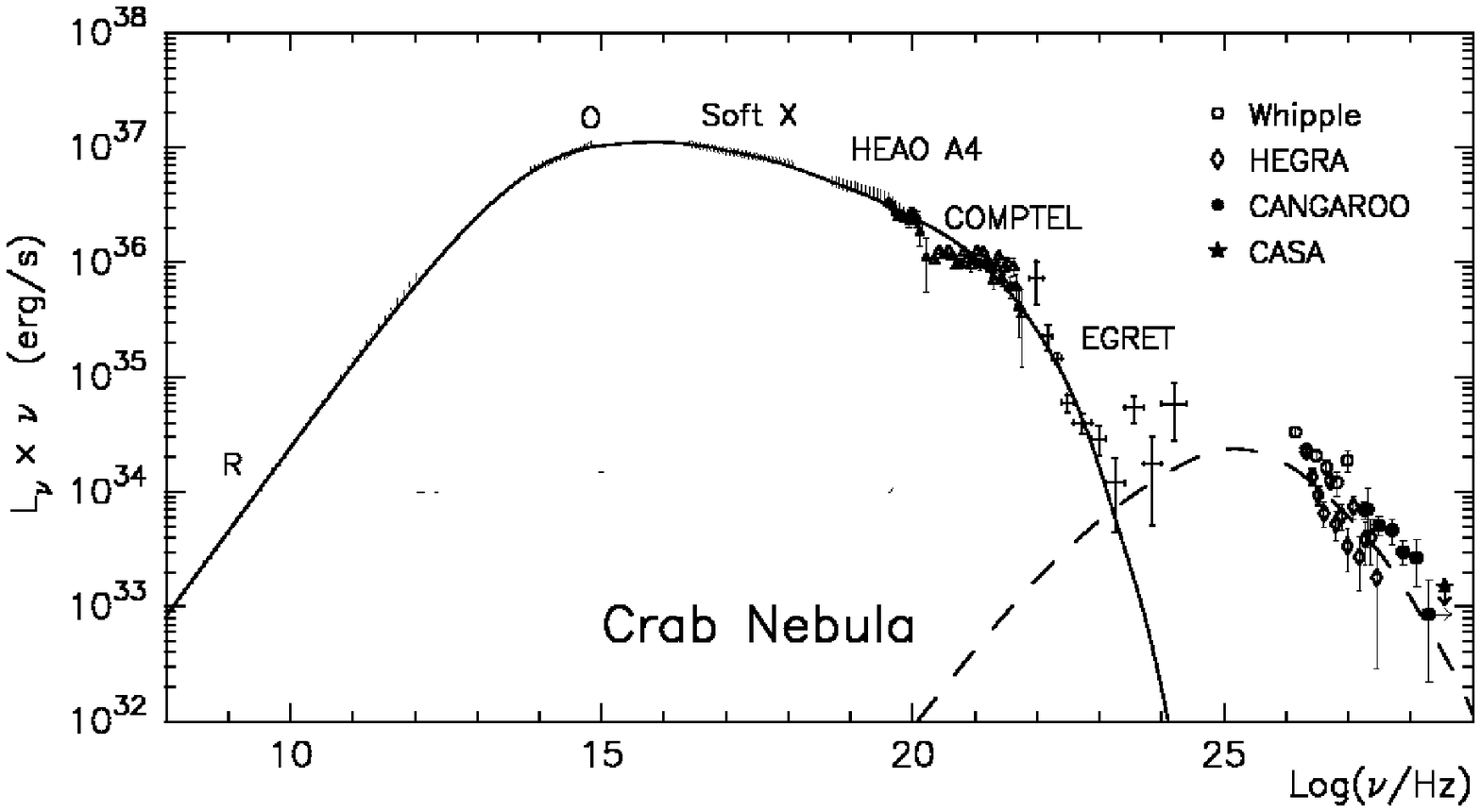}
\caption{Multiwavelength spectrum of the Crab nebula. The lines
represent a two component model; the solid line is synchrotron
emission and the dashed line is IC emission. Adapted
from Atoyan and Aharonian (1996).}
\label{fig:aharoniancrab}
\end{figure}

The spectral energy distribution of the Crab Nebula, the best studied
plerion, is shown in Fig.~\ref{fig:aharoniancrab}. There appear to
be two components of the emission:
synchrotron emission is believed to responsible for low-energy \grays\
down to radio, while IC scattering off CMB, dust far-IR, and
synchrotron photons, is the most probable mechanism of high-energy
\gray\ emission \cite{atoyan96}. The energetic electrons responsible
for this emission are provided by the pulsar wind or the wind
termination shock.  Another possibility which cannot yet be excluded
is that some part of the \gray\ emission may be due to $\pi^0$-decay
\grays.

Observations in the sub-TeV range will probe the emission mechanisms in
plerions; allowing us to derive the nature of accelerated particles
(leptons, hadrons), their spectra, and the intensity of the magnetic
fields in the nebulae.

\subsection{Blazars and radio galaxies}

One of the surprises in AGN astrophysics over the past 10 years has
been the detection of blazars at very high \gray\ energies. Most of
the high latitude sources detected by EGRET were blazars, which
are AGN identified at other wavelengths to be either flat spectrum
radio quasars (FSRQs) or BL Lacs. EGRET detected about 66 blazars of
which 46 were FSRQs and 17 were BL Lacs \cite{hartman99}, at higher
energies ($>$300 GeV) several BL Lacs have been detected by ground
based \gray\ instruments \cite{punch92,quinn96}.

These objects are highly variable at all frequencies. High energy
\gray\ flares have been observed on timescales of minutes to months.
In fact, most of these AGN were only detectable by EGRET during
\gray\ flares. Several multiwavelength campaigns have revealed that
the \gray\ flares are frequently correlated with variability at longer
wavelengths \cite{wehrle98,buckley96}. The \gray\ spectrum of
blazars measured by EGRET is adequately fit by a simple power-law,
although the data do not exclude more complicated shapes. The average
spectral index in the EGRET energy band is about --2.2, with no
evidence for a spectral cutoff for energies below 10
GeV \cite{mukherjee99}. 

The broadband spectral distributions are characterized by two
components, one extending from radio to X-rays, interpreted as being
due to synchrotron radiation, and another higher energy component. The
origin of the high energy component is still a matter of considerable
debate.  Currently favored models include (a) electron IC
scattering of synchrotron photons from within the jet or photons
produced outside the jet, (b) proton initiated cascades, or (c) proton
and muon synchrotron radiation. Simultaneous broadband observations of
blazars, both in flaring and quiescent states, provide excellent
tests of emission models \cite{buckley96,vonmontigny95,sikora01}.

In the most leptonic models, leptons accelerated by the central engine
emit \grays\ via IC scattering of background photons
\cite[and references therein]{katarzynski01}.  The background photons
may be internal or external.  Internal photons may be the synchrotron
photons emitted by earlier generation of electrons
(synchrotron-self-Compton models), while external can be CMB photons
or photons reprocessed by a surrounding matter, e.g., gas clouds or
the accretion torus (external IC models).

In the hadronic scenario, it is assumed that accelerated protons
produce $e^\pm$ pairs in $pp$- or $p\gamma$-interactions. The pairs
produce synchrotron or IC photons, which we see or, if
the optical depth near the source is high, produce new pairs via
$\gamma\gamma$ interactions and so on.  To start the electron-photon
cascade in this scenario there needs to be either a suitable gas
target for accelerated nucleons, or the nucleons need to be
accelerated to energies high enough to create pairs on background photons.

\begin{figure}[tb!]
\vskip 3.35in
\includegraphics{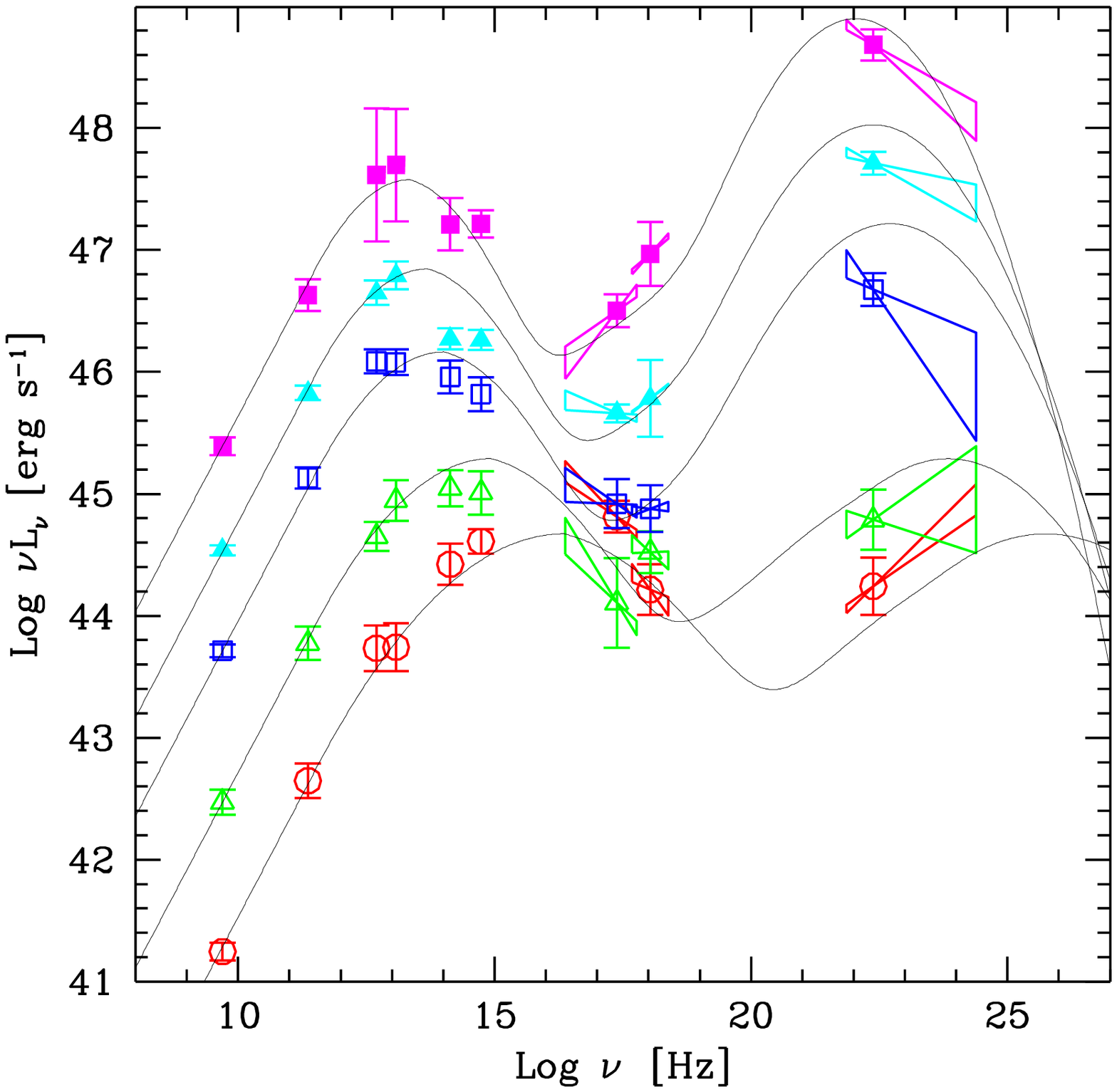}
\caption{The points represent the average SEDs derived by
\cite{fossati98}.  BL Lacs and FSRQs belonging to complete samples
have been divided into bins according only to the radio
luminosity. The solid lines are the parameterisations of the spectra
proposed in Donato et al.~(2001).}
\label{fig:fossati}
\end{figure}

There may be systematic differences in several properties of blazars
related to the location of the synchrotron peak, illustrated in
Fig.~\ref{fig:fossati}.  As the synchrotron peak moves to lower
frequencies; the ratio of the powers of the high to the low energy
spectral components increases, the total power output of the blazar
increases and emission lines become stronger, resulting in the
sequence FSRQs, low-energy peaked BL Lacs and finally high-energy
peaked BL Lacs. These differences have been interpreted as resulting
from orientation \cite{ghisellini89,urry95} or as intrinsic changes in
physical parameters in the jet or jet
environment \cite{padovani95,ghisellini98}. However, it is difficult
to draw many conclusions about general properties of blazars from
population studies at \gray\ energies. The sample of \gray\ blazars
suffers from selection effects, GeV blazars were generally only seen
during flares and only a few blazars have been detected at TeV
energies. Some recent blazar surveys suggest that the apparent blazar
sequence may be at least partially due to selection
effects. Population studies by GLAST will help the study of blazars as
a class by providing an unbiased set of data.

The peak sensitivity of the LAT lies at somewhat higher energies than
that of EGRET. Thus it is likely that the LAT will identify an
additional population of blazars intermediate in properties between
the EGRET detected and the TeV blazars. These are likely to include a
much higher fraction of BL Lacs than were present in the EGRET
sample. Studies of the high energy $>$10 GeV photons from blazars in
the EGRET data suggest that they may be predominatly from BL Lacs,
this work provides a taste of what might be still to come from GLAST.
The actual number of blazars that will be detected by the LAT depends
on the luminosity function and cosmological evolution of these objects,
which, as yet, are unknown. Estimates of the number of blazars that will
be detected by the LAT range from about 2500 to about 10000
objects. It is clear that the number of known \gray\ blazars will
dramatically increase allowing for a much more detailed study of the
underlying physical processes responsible for the blazar sequence.

The large field of view of the LAT will be of enormous importance for
blazar observations. Most of the EGRET blazars were detected only when
they flared. Since GLAST monitors (in the scanning mode) the entire sky on
all timescales greater than an hour, all AGN anywhere in the sky that
flare above the LAT detection threshold will be detected. This will
likely lead to a dramatic increase in the number of known blazars and
will require prompt follow up at many wavelengths to properly study
and catagorise these sources.

\begin{figure}[tb!]
\vskip 3.3in
\includegraphics{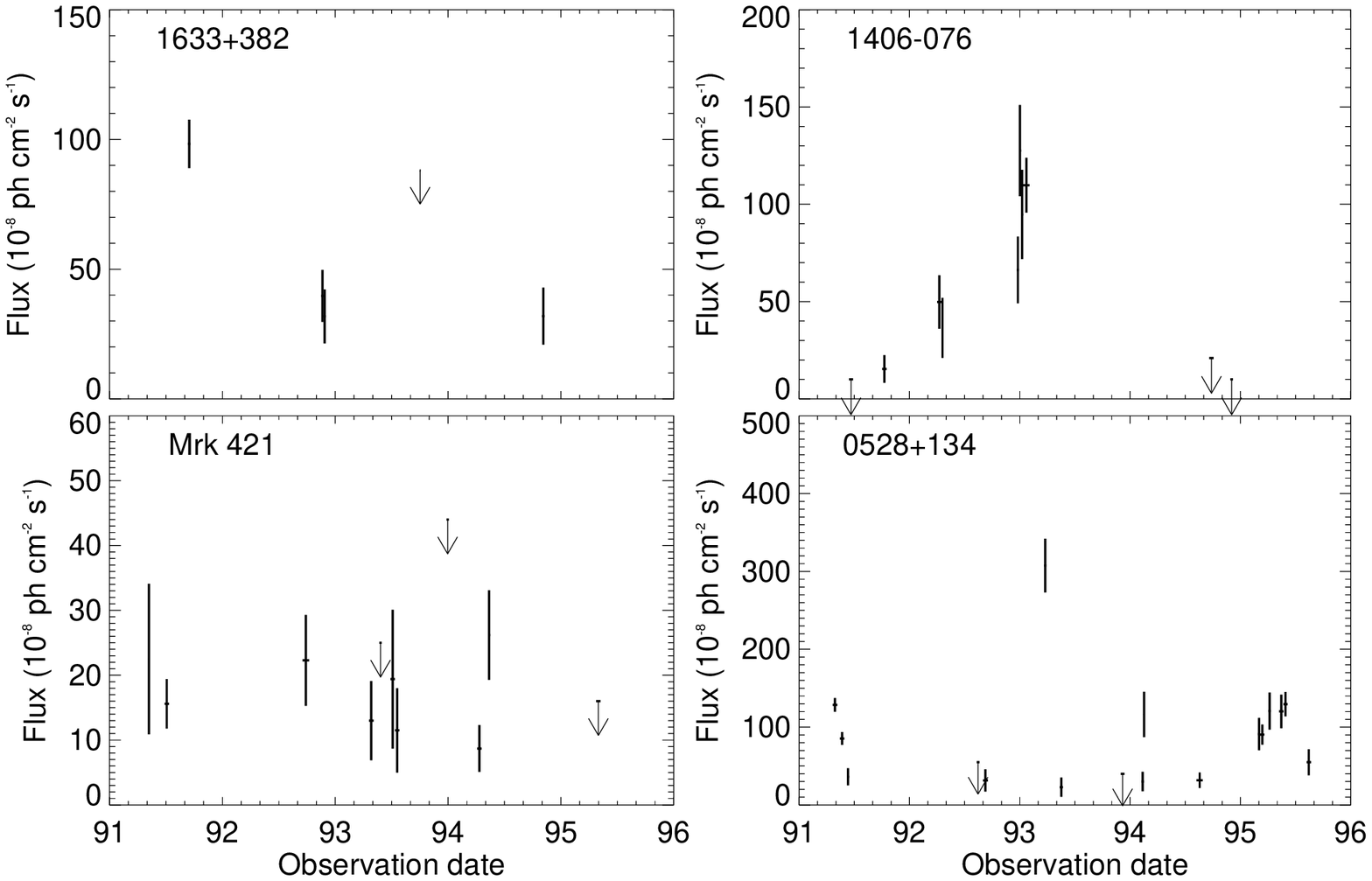}
\caption{Flux history of 4 blazars \cite{mukherjee99}: PKS 1633+382, PKS 1406--076, 
Mrk~421, and PKS 0528+134.}
\label{fig:mukherjee}
\end{figure}

Also of crucial importance is the ability of the LAT to generate
detailed, complete lighcurves on all AGN in the sky on timescales down
to hours. This will be a dramatic improvement over what was possible
with EGRET. This is illustrated by Fig.~\ref{fig:mukherjee} which
shows the flux history of 4 blazars \cite{mukherjee99}. The
horizontal error bars indicate the duration of the observation. The
limitations of these observations are immediately apparent. Flux
variabilty can only be studied either on short timescales within an
observation period, or on very long timescales with poor sampling
between viewing periods. The large field of view of the LAT will allow
continuous observations with increased sensitivity making possible
studies on shorter timescales and of less intense flares.

Evidence for very high energy \gray\ emission has been reported for
two radio galaxies. An EGRET source location is consistent with the location of
Centaurus A \cite{sreekumar99}, the brightest AGN in the hard X-ray
sky and one of the nearest. The HEGRA collaboration have reported
emission from M87, another nearby
radio-galaxy \cite{aharonian03}. Both M87 and Cen A are considered
to be misaligned blazars, i.e. radio-galaxies aligned such that the
relativistic jets are at large inclination angles to our line of
sight \cite{morganti92,bicknell96}. Both of these detections are at
somewhat marginal significance, and represent a significantly lower
\gray\ flux than most blazars. However, these detections raise
exciting possibilities with a more sensitive instrument like the LAT
as the space density of radio galaxies are $\sim$10$^3$ times that of
blazars. Confirmation of \gray\ emission from radio galaxies and
detection of a larger sample at a variety of inclination angles would
provide valuable information about the parent population of blazars
and the energy dependence of the beaming cones of blazars. Once the
beaming cone is known, statistical studies can establish the paternity
of BL Lacs and quasars to radio-loud Faranoff-Riley I and II
galaxies \cite{urry95}.

\subsection{Gamma-ray bursts}

Gamma-ray bursts are extremely intense, but short-lived sources
of \grays. Observations are generally made of two phases of a burst:
the prompt phase, which is a short (milliseconds to thousands of
seconds) period of very intense \gray\ emission and the afterglow
phase which is observed at radio, optical, X-ray and \gray\ wavebands
and decays on timescales of hours to days. High energy \gray\
observations of both phases are extremely important in
understanding this extraordinary phenomenon.

EGRET detected four GRBs above 100 MeV, which were the brightest that
occurred within the field of view of its spark chamber. EGRET also
detected 30 GRBs above 1 MeV in its calorimeter. The high energy
spectra are consistent with a power-law with no evidence for a cutoff
below energies of 10 GeV. In one noteworthy burst, GRB 940217, EGRET
detected high-energy emission persisting for about 5000 seconds beyond
burst cessation at hard X-ray energies \cite{hurley94}; this
high-energy \gray\ afterglow contained a significant fraction of
the total burst fluence. A few other bright GRBs observed by EGRET
also show indications of longer duration emission \cite{dingus97}.
A recent analysis of archival data from the EGRET calorimeter has
found a multi-MeV spectral component in the prompt phase of GRB 941017,
that is distinct from the lower energy component. This observation is 
difficult to explain within the standard synchrotron model thus indicating
the existance of new phenomena \cite{gonzalez03}.

One of the most widely accepted models of the \gray\ burst phenomenon
postulates that they are powered by a relativistically expanding
fireball. Electrons accelerated at shocks produced by colliding shells
of material inside the fireball (known as internal shocks) produce the
radiation during the prompt phase via synchrotron emission. The
afterglow is due to emission from non-thermal particles when the
external shock formed when the fireball blastwave sweeps up the
external medium. The high energy photons are (theoretically) the most
difficult to produce and are easily lost due to conversion to
e$^+$e$^-$ pairs. They are thus of particular importance in constraining
\gray\ burst physics.

There are several important questions about high energy emission from
\gray\ bursts which remain to be answered. How high in energy does the
emission extend? The measurement of the highest energy photons allows
the determination of a constraint on the bulk Lorentz factor of the
expansion. Pair production with lower energy photons will attenuate
the high energy \grays\ unless the bulk Lorentz factor is large enough
so that the \grays\ do not have sufficient energy to produce pairs
in the rest frame of the fireball. It is important to establish
whether there is, in general, a second higher energy component of
emission in either the prompt or afterglow phases of a \gray\ burst,
understanding the nature of such emission will provide important
information about the physical conditions of the emission
region \cite{zhang01,pilla98,dermer99}.

One of the most important improvements of the LAT compared to EGRET is
the very much reduced deadtime.  The EGRET deadtime was about 110 ms
per photon, which is comparable to the \gray\ pulse widths at lower
energies where the time profiles are well measured. The deadtime of
the LAT will be $<$100 $\mu$s, so \gray\ burst observations will not
be limited by the deadtime. This, combined with the large collection area,
will allow the LAT to detect bursts to much lower intensities and to
study the lightcurves on much finer structures than has previously
been possible. The higher energy reach of the LAT will allow a
systematic study of \gray\ bursts above 1 GeV.  The LAT's field of
view is more than four times that of EGRET which will result in many
more bursts being detected. With some reasonable assumptions of the
flux and spectral index distributions it has been predicted that the
LAT will see $>$200 bursts per year.

\subsection{Extragalactic diffuse emission} \label{egb}

The extragalactic diffuse \gray\ emission is of particular interest as
the bulk of these photons suffer little or no attenuation during the
propagation from their site of origin. It is thus a superposition of
all unresolved sources of high energy \gray\ emission in the
Universe. The isotropic, presumably extragalactic component of the
diffuse \gray\ flux was first discovered by the SAS-2 satellite and
confirmed by EGRET \cite{thompson82,sreekumar98}.

The spectrum of extragalactic diffuse \gray\ emission (EGB) is the
most difficult component of the diffuse emission to extract. It
depends strongly on the adopted model of the Galactic background which
itself is not yet firmly established.  It is not correct to assume
that the isotropic component is entirely extragalactic, because even
in the pole direction it is comparable to the Galactic contribution
from IC scattering of photons from the Galactic plane and CMB photons
off the halo electrons \cite{moskalenko00}.  The modelled
emission depends on the size of the halo, the electron spectrum there,
and the spectrum of low-energy background photons.

\begin{figure}[tb!]
\vskip 3.6in
\includegraphics{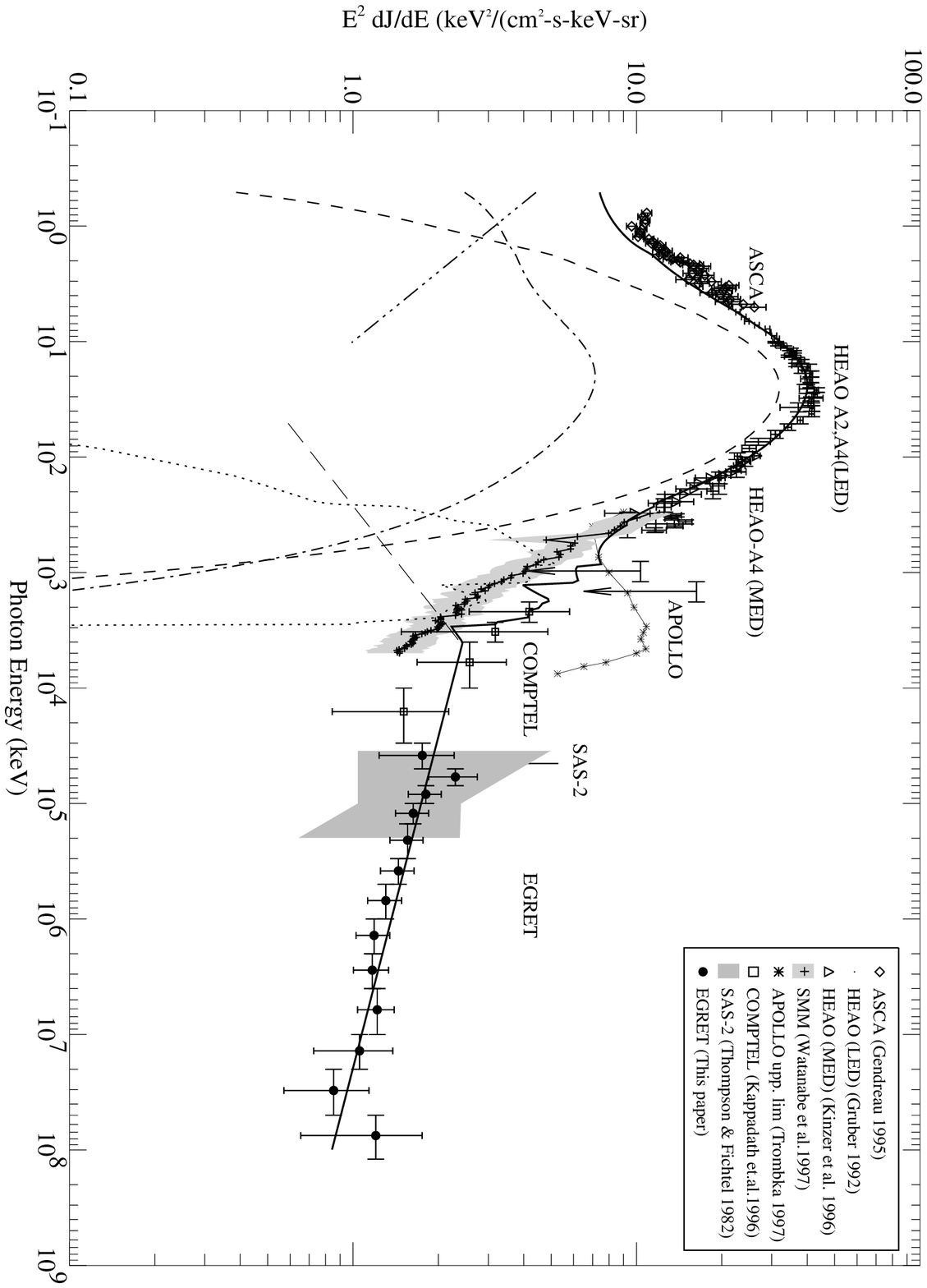}
\caption{Multiwavelength spectrum of the diffuse extragalactic emission from
X-rays to \grays\ taken from Sreekumar et al.~(1998). The dot-dashed and dashed 
lines are estimates of the contributions of Seyfert I and Seyfert II galaxies
respectively. The triple dot-dashed line represents the steep-spectrum quasar
contribution and the dotted line is for type Ia supernovae. The long dashed
line is an estimate of the contribution from blazars. The thick solid line
is the sum of all components. }
\label{fig:sreekumar}
\end{figure}

\begin{figure}[tb!]
\vskip 3.6in
\includegraphics{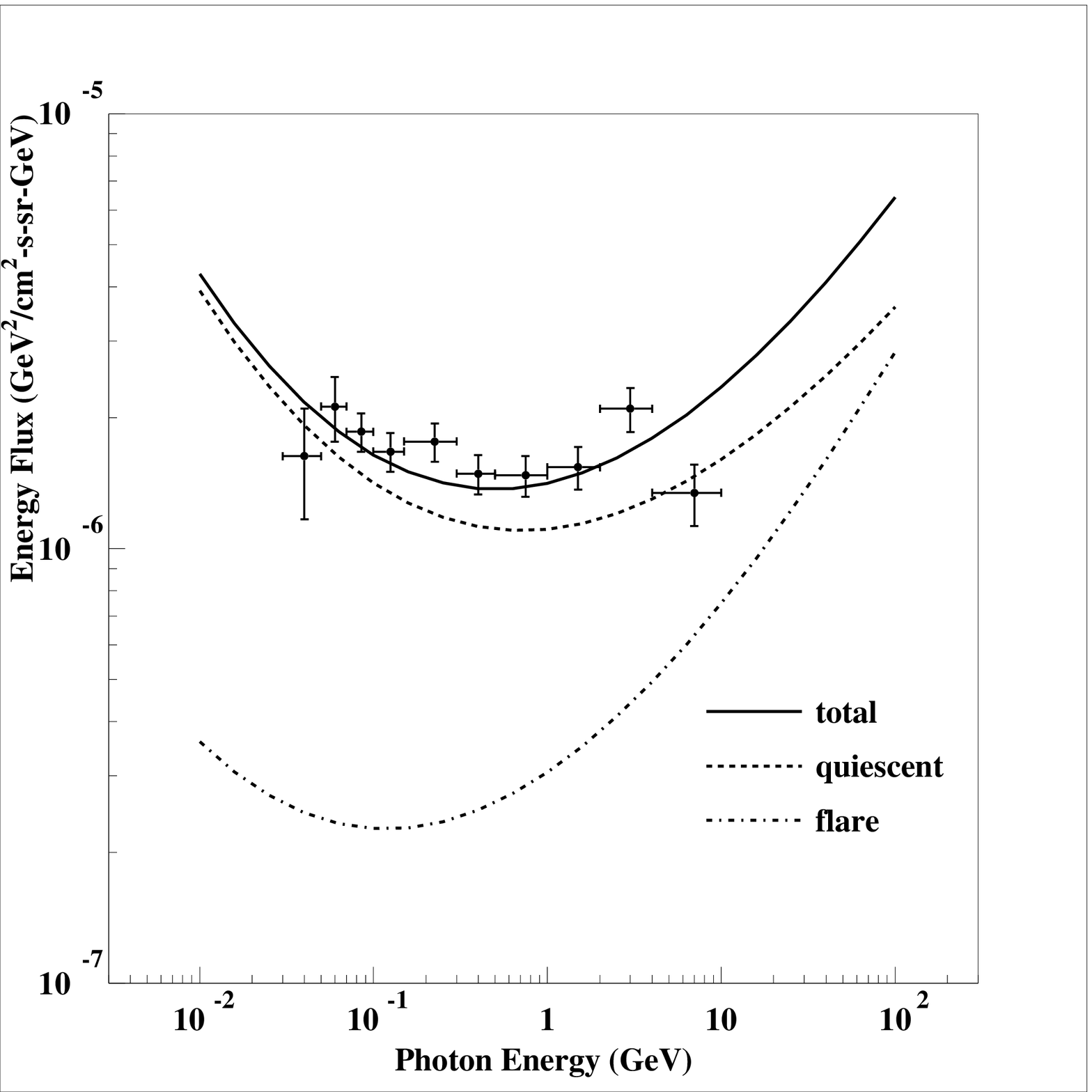}
\caption{Predicted background spectra from unresolved blazars. The lower middle
and upper curves show the contribution to the EGRB from flaring sources, from
quiescent sources and their total \cite{stecker96}.}
\label{fig:stecker}
\end{figure}

An extensive analysis has been undertaken \cite{sreekumar98} to derive
the spectrum of EGB based on EGRET data which is shown in
Fig.~\ref{fig:sreekumar}.  The distribution of
modelled-Galactic-diffuse-emission along with the catalog of point
sources were carefully subtracted from total-diffuse-emission to
determine the EGB as an extrapolation to zero Galactic contribution.
The spectral index that was obtained by this analysis, $-2.10\pm0.03$,
appears to be close to that of \gray\ blazars.

The origin of the isotropic background is uncertain, several
possibilities have been proposed including barion-antibarion
annihilation in the early Universe \cite{stecker71}, evaporation of
primordial black holes \cite{page76} and annihilation of exotic
particles \cite{silk84,rudaz91} in addition to the contribution
from unresolved extragalactic sources such as AGN, \gray\ bursts and
galaxy clusters. Thus observations of the EGB can provide constraints
on the existance of these phenomena and on their cosmological
evolution.
 
Since most \emph{identified} sources of extragalactic \grays\ are AGN,
the \emph{unresolved} AGN are likely to contribute to the EGB.
There
have been several analyses of the expected level of this contribution.
Stecker and Salamon (1996) assumed that the \gray\ luminosity of
blazars was proportional to their radio luminosity and used the
measured radio luminosities to derive a luminosity function for \gray\
AGN; when integrated, this luminosity function indicates that
unresolved blazars could contribute 100\% of the extragalactic diffuse
emission (Fig.~\ref{fig:stecker}). 
A direct determination of the blazar luminosity function
from EGRET data leads to an estimate that the blazar contribution to
the EGB is around 25\% \cite{chiang98}. In a
more recent analysis, the \gray\ luminosity
function \cite{mucke00} is derived from a \gray\ emission model
combined with an expected dependence of the AGN luminosity
from orientation of the jet with respect to the line of sight.
The results of this analysis suggest that
between 20--40\% of the EGB emission can be explained by unresolved
AGN for an AGN luminosity function with a break at redshifts greater
than 3 or even as much as 40--80\% if the redshift cutoff occurs at
$z=5$. 

\begin{figure}[tb!]
\vskip 3.6in
\includegraphics{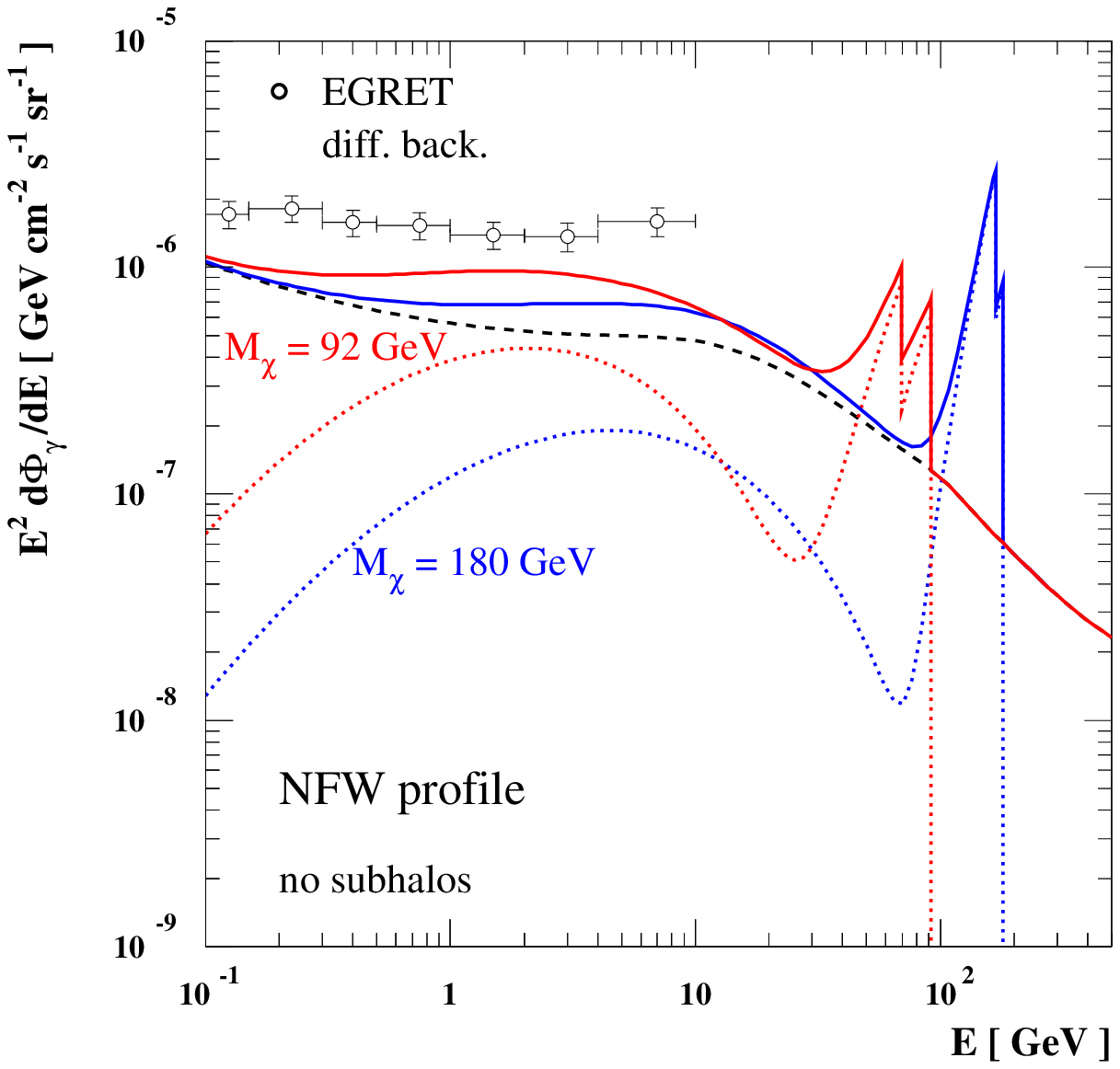}
\caption{Extragalactic \gray\ flux (multiplied by E$^2$) for two sample
thermal relic neutralinos (dotted curves), summed with the unresolved 
blazar background expected for GLAST (dashed curve) \cite{ullio02}.}
\label{fig:ullio}
\end{figure}

Observations with the LAT will improve these limits in two ways:
(i) studies of a larger population of AGN will allow the better
determination of the blazar luminosity function and thus a more
reliable estimation of the contribution from unresolved blazars, and
(ii) the improved sensitivity and angular resolution means that
many more AGN will be detected as resolved sources, thus lowering the
residual isotropic background from unresolved sources. This is
particularly interesting as it raises the exciting possibility of
detecting the spectral signatures of a more speculative truly diffuse
component. An example of such a spectral feature from the annihilation
of relic neutralinos is shown in
Fig.~\ref{fig:ullio} \cite{ullio02}.

\subsection{Supernova remnants and cosmic-ray acceleration sites}

One of the long-standing problems in astrophysics is the origin of
cosmic rays. EGRET observations of the Small and Large Magellanic
Clouds \cite{sreekumar92} have shown that cosmic rays are likely to
be Galactic in origin.  Because of their energetics, supernova remnants 
(SNR) are thought
to be the main sources of cosmic rays, at least, up to the ``knee'' in
the cosmic-ray spectrum ($\sim$$3\times10^{15}$ eV).  Particles are believed to be
accelerated by the shock waves of the SNR in the first-order Fermi
diffusive acceleration mechanism, which should work for both
leptons and nucleons.  The resulting spectrum of accelerated particles is
expected to approximate a power-law.

\begin{figure}[tb!]
\vskip 3.2in
\includegraphics{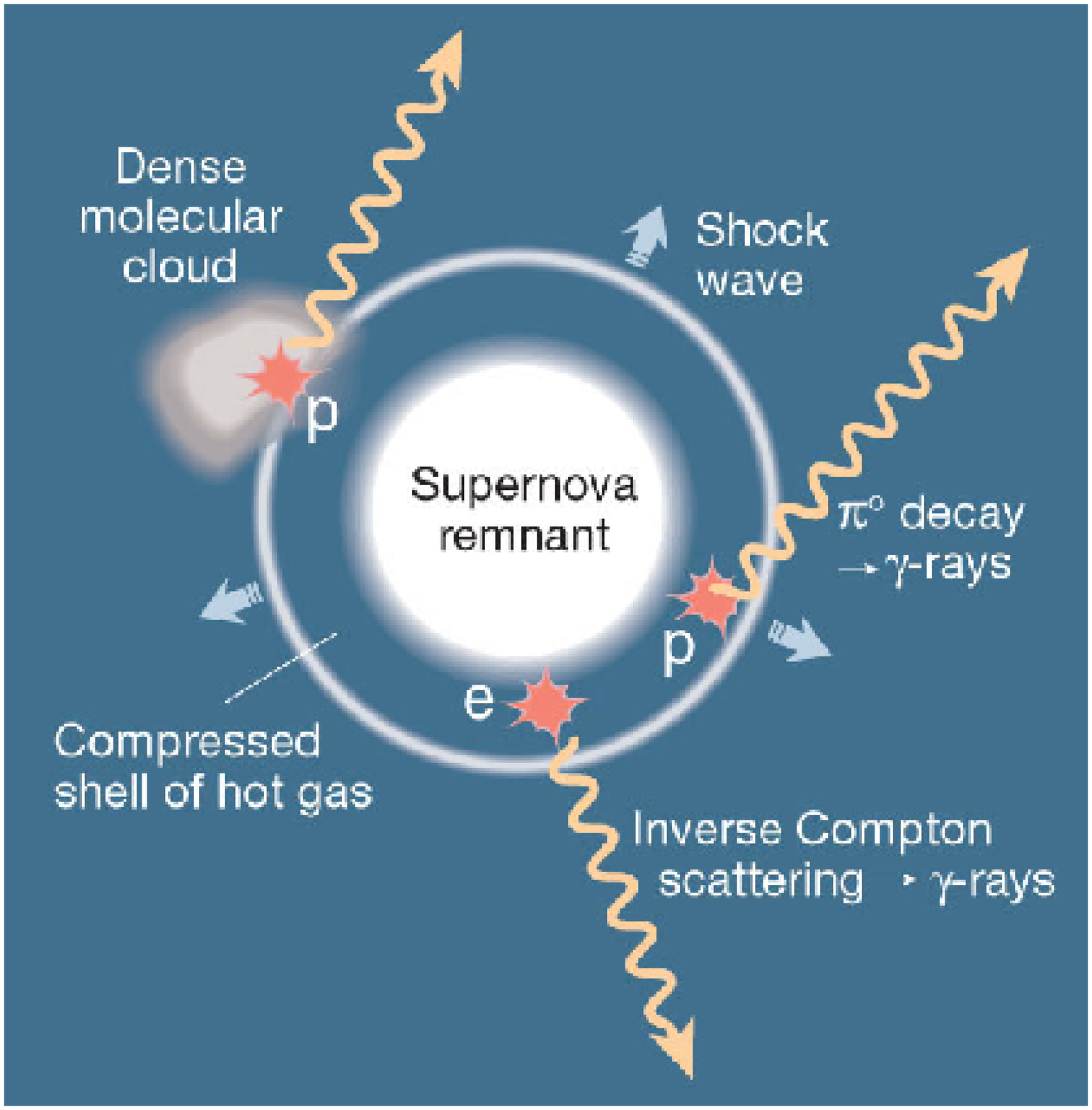}
\caption{Shock waves from SNR can produce high-energy \grays\ in two ways: IC
scattering of CMB photons off relativistic electrons, 
and $\pi^{0}$-decay from interactions of relativistic protons in the
ISM.
Adapted from Aharonian~(2002).}
\label{fig:aharonian}
\end{figure}

SNR can produce high-energy \grays\ of nucleonic and/or leptonic origin (see
Fig.~\ref{fig:aharonian}), however the spectral shapes of each are
distinctly different \cite{gaisser98}.  The bremsstrahlung spectrum
is steep and thus dominates at lower energies, while the IC
spectrum is flat and contributes to higher energies.  \grays\
resulting from $\pi^0$ decays display a bump in the spectrum which
contributes to the intermediate energies (GeV-sub-TeV).  This is a
general picture, which may be not correct in details.  Because of
large energy losses of leptons via synchrotron and IC scattering the
spectral index of accelerated leptons may be steeper than that of
nucleons, thus the regions of dominance of different mechanisms can
be, in reality, different.

Until recently, however, there was no direct evidence of particle
acceleration in SNR. Strong non-thermal X-ray emission observed by
ASCA from SN 1006 \cite{koyama95}, by ROSAT from RX J1713.7-3946
\cite{pa96}, by RXTE from Cas A \cite{allen97}, and several others
have been interpreted as synchrotron emission of electrons accelerated
up to $\sim$100 TeV. Detections of GeV \grays\ from the regions near
several SNR have been reported by EGRET \cite{esposito96}.  A
number of unidentified EGRET sources have been statistically
attributed to SNR \cite{sturner95}.  Recent evidence of TeV emission
from SN 1006 \cite{tanimori98b} and RX J1713.7-3946
\cite{muraishi00} has been reported by the CANGAROO collaboration,
and a detection of Cas A has been reported by the HEGRA collaboration
\cite{aharonian01}. These detections represent the first
\emph{direct} evidence of particles, most likely leptons, up to
$\sim$100 TeV.  The observed TeV fluxes are consistent with being
IC emission and the magnetic field of $\sim$$10\mu$G can
be derived from observations of the synchrotron emission.

\begin{figure}[tb!]
\vskip 1.1in
\includegraphics{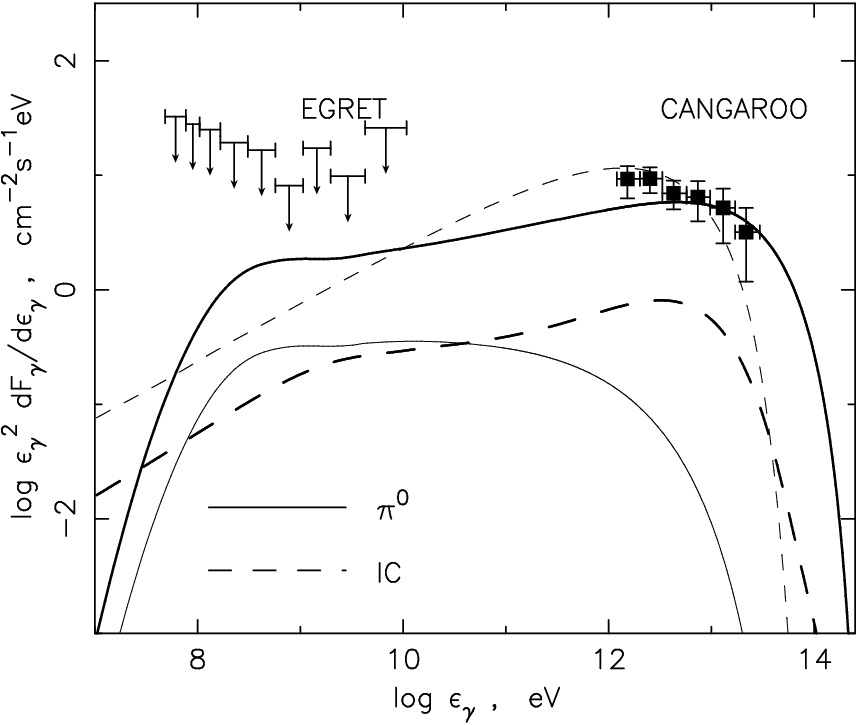}
\narrowcaption{SN 1006.
Differential $\pi^0$-decay (solid lines) and IC (dashed lines) for
efficient (thick lines) and inefficient (thin lines) proton acceleration
cases \cite{berezhko02}.}
\label{fig:berezhko}
\end{figure}

\begin{figure}[!tb]
\vskip 0in
\includegraphics{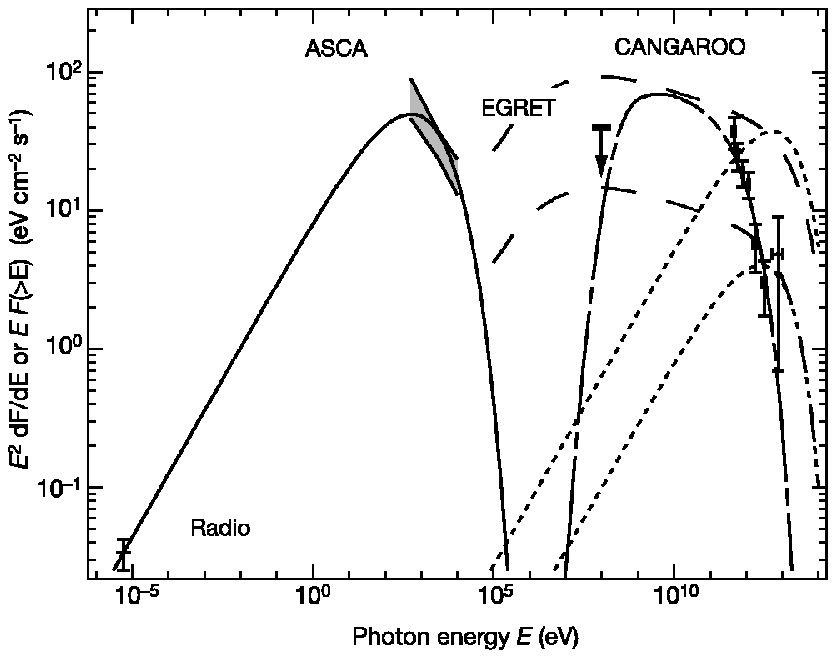}
\narrowcaption{Multi-band emission from RX J1713.7--3946, and emission models.
Model calculations: solid line -- synchrotron emission, dotted lines --
IC scattering, dashed lines -- bremsstrahlung, short-long dahed line -- $\pi^0$-decay.
The IC and bremsstrahlung are shown for $3\mu$G (upper curves),
and $10\mu$G (lower curves).
Adapted from Enomoto et al.~(2002).
\label{fig:enomoto} }
\end{figure}

While the high energy lepton population in remnants clearly manifests
itself via synchrotron X-rays, acceleration of nucleons to high
energies has yet to be observed. The estimated flux of \grays\ from
$\pi^0$-decay at TeV energies is small; to detect it one needs an
instrument capable of observations at sub-TeV energies with good
angular resolution.  There have been some arguments that the TeV
\grays\ from SN 1006 \cite{berezhko02}, illustrated in
Fig.~\ref{fig:berezhko}, and RX J1713.7--3946 \cite{enomoto02} in
Fig.~\ref{fig:enomoto} are
of hadronic origin, however the arguments and observations for this
are not yet conclusive \cite{reimer02,butt02}.

Ultimate evidence of hadron acceleration would be $\pi^0$-decay
\grays\ from a molecular cloud in the vicinity of a SNR
\cite{aharonian94,aharonian96}.  An unidentified EGRET source 3EG
J1714--3857 has been interpreted \cite{butt01} as being a signature
of $\pi^0$-decay \grays\ from nucleonic interactions in dense
molecular clouds found in the vicinity of the TeV source RX
J1713.7--3946.  Another possible signature of $\pi^0$-decay \grays\ is
the unidentified TeV source which has been found in the vicinity of
Cygnus OB2 association \cite{aharonian02}.

The measurements of the \gray\ spectrum of a number of SNR by GLAST
will help to solve several long-standing problems.  The origin of
cosmic-ray remains uncertain, the LAT should be able to address
whether SNR are accelerating protons and, if so, to which energy. The
LAT will also be able to study the energy output of SNR in high
energy particles and probe electron/proton ratio of the accelerated
particles. The understanding of the dominant mechanisms producing
\grays\ in SNR will provide a wealth of information on the physical
conditions near SNR, such as the gas density, the background photon
field, and the magnetic field. The derivation of the spectra of
accelerated particles will be very important in better understanding
theories of shock acceleration.
More detailed discussion of particular SNR and extensive references
can be found in Chapter 5 and in Torres et al.~(2002).

\subsection{Galaxy clusters}

Galaxy clusters have yet to be observed in \grays, but their
significant content of nonthermal particles implies that their \gray\
emission should be detectable by the next generation of \gray\
telescopes.  A signature of the presence of relativistic electrons in
the intracluster medium is the detection of a diffuse cluster-scale
radio emission from a number of these objects \cite[and references
therein]{giovannini99}.  The intracluster magnetic field estimates
range from $0.1\mu$G to $1\mu$G leaving a large window for
speculations of the total energy content of relativistic electrons
(and protons) there.

The non-thermal electrons in clusters are believed to be produced in
several ways. Among the mechanisms discussed are the acceleration by
merging shocks and injection of relativistic particles by active
galaxies which may be members of the cluster.  A significant fraction
of nonthermal electrons can be produced by high energy protons in
$pp$-interactions via production and decay of charged pions and from
annihilation of supersymmetrical particles.
Because of large energy losses electrons are to be
accelerated in a relatively short time scale of a few Byr,
while high-energy protons can be accumulated on cosmological time scale.

Galaxy clusters are among the primary candidates to be associated with
unidentified EGRET sources.  Fifty out of 170 unidentified sources in
the Third EGRET catalogue \cite{hartman99} are found at intermediate
and high Galactic latitudes, $|b|\ge20^\circ$, and are likely to be
extragalactic.  The low flux variability of these sources implies that
they can not be attributed to a highly variable AGN
population. Spatial correlation with galaxy clusters and correlation
between 1.4 GHz radio flux from the clusters and \gray\ flux supports
this hypothesis \cite{cola02}.

\begin{figure}[!tb]
\vskip 2.5in
\includegraphics{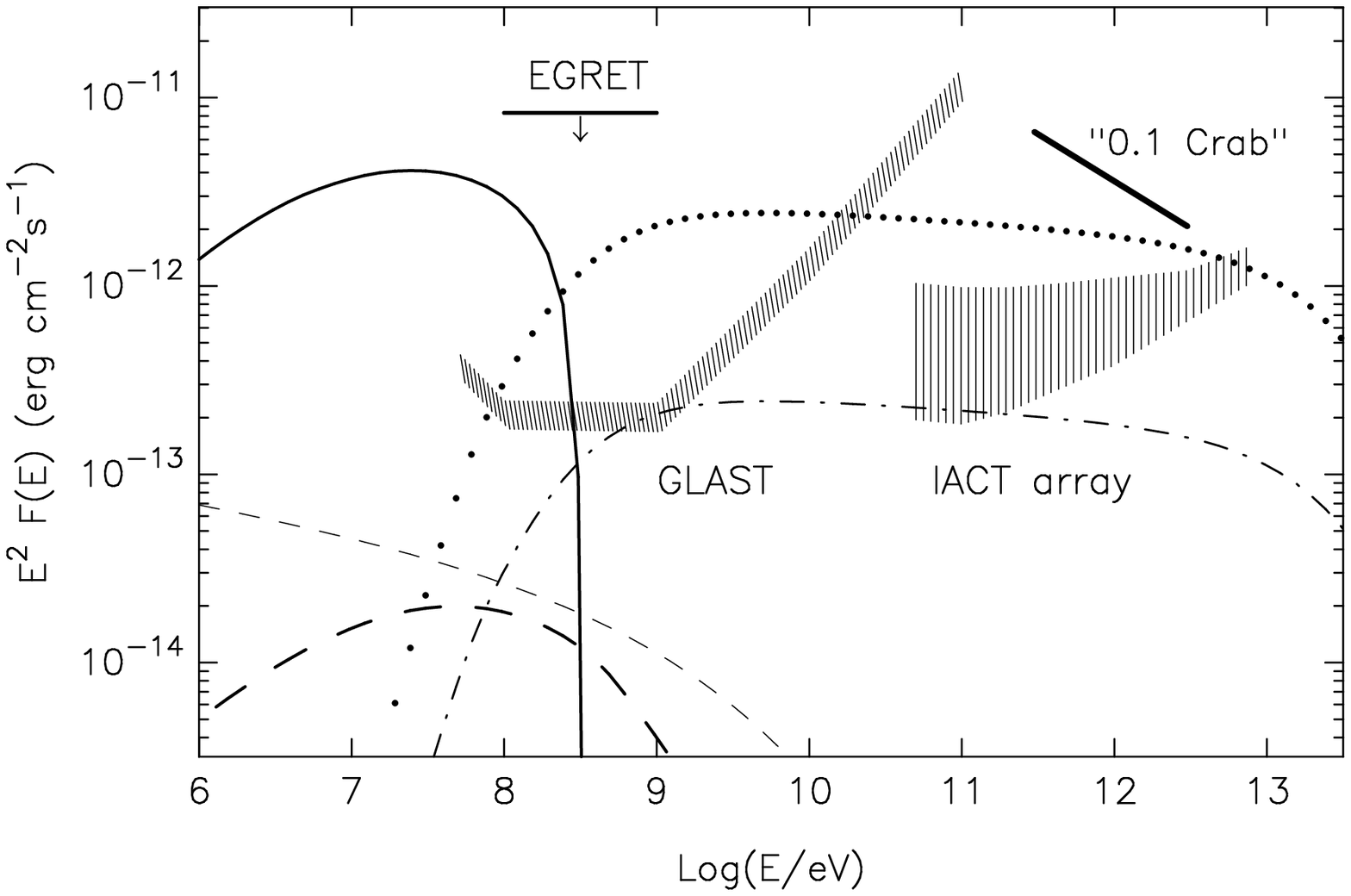}
\caption{\gray\ flux expected from the Coma cluster. The solid curve 
shows the bremsstrahlung flux produced by the relic population of electrons
(derived from the UV flux). Heavy dashed and thin dashed show the 
bremsstrahlung and IC fluxes produced by radio electrons ($B=1\ \mu$G).
Dotted and dash-dotted curves correspond to the $\pi^0$-decay produced by
protons (index $-2.1$)
with total energy $3\times10^{62}$ and $3\times10^{61}$ erg, 
correspondingly \cite{atoyan00}.}\label{fig:atoyan00}
\end{figure}

The \gray\ fluxes from electrons (bremsstrahlung, IC scattering) and
protons ($\pi^0$-decay) have been predicted for some clusters using
the radio, EUV, and X-ray observations.  In the case of the Coma
Cluster, EUV and X-ray observations show excesses 
compared to single-temperature plasma models \cite{fusco99,rephaeli99}. If
interpreted in terms of IC scattering of CMB photons by electrons of some 100
MeV \cite{ensslin98,sarazin98}, it yields the intracluster magnetic
field of $0.1\mu$G \cite{atoyan00} in contrast to $\sim$$1\mu$G derived
from Faraday rotation measurements.  A magnetic field of $\sim$$1\mu$G
would produce synchrotron flux one order of magnitude higher than
observed while falling short of explaining the EUV and X-ray fluxes
due larger electron energy losses (Fig.~\ref{fig:atoyan00}).

\begin{figure}[tb!]
\vskip 3.3in 
\includegraphics{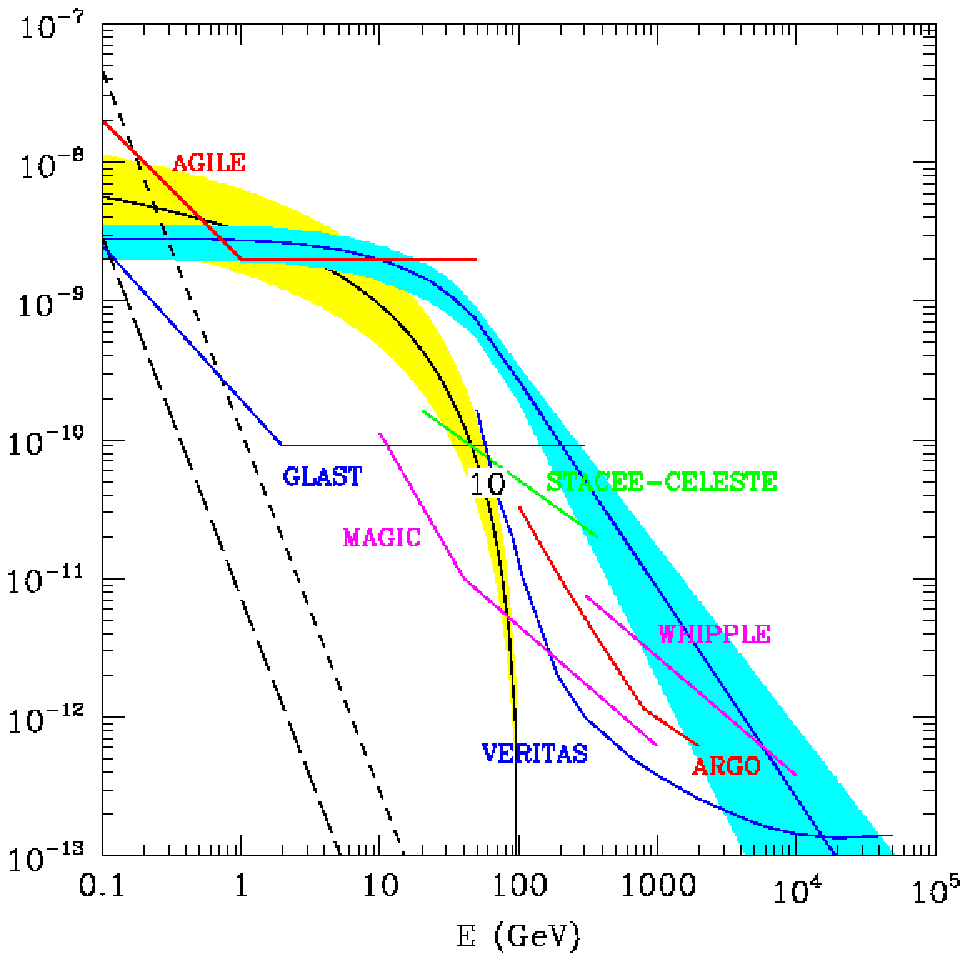} 
\caption{Coma
cluster. Predictions for the \gray\ flux, $F_\gamma(>100 {\rm MeV})$
cm$^{-2}$ s$^{-1}$, expected for a Coma-like cluster are compared with
the sensitivity of the next generation \gray\ experiments. Lines are
coded as following: short-dash -- electron bremsstrahlung for $B=0.3\
\mu$G, long-dash -- electron bremsstrahlung for $B=1\ \mu$G, black
curve -- $\pi^0$-decay from $pp$-interactions and associated
uncertainties (dark shaded region), black solid line -- $\pi^0$-decay from
neutralino annihilations and associated uncertainties (light shaded
region) \cite{cola02}.} \label{fig:cola02}
\end{figure}

GLAST sensivity will allow it to detect for the first time \grays\
from the galaxy clusters (Fig.~\ref{fig:cola02}).  
Such observations will help to derive the
relative importance of \gray\ emission mechanisms and study physical
conditions in the very-large-scale regions of the Universe.
It would uncover mechanisms of particle acceleration in clusters.

\section{Conclusions}

One of the surprising results of observations above 100 MeV is the
small number of classes of \gray\ sources 
(Table~\ref{tab:sources}).  At energies between
100 MeV and 30 GeV the only confirmed sources are pulsars/plerions,
blazars, GRB, the sun and the Large Magellanic Cloud. In addition,
based on spectral arguments, the nearby radio galaxy Cen A is believed
to be associated with one of the sources in the EGRET 3rd catalog.
At energies above 300 GeV the number of sources is very much smaller.
At these energies several blazars and plerions have been detected by
more than one instrument. In addition, there have been reported
detections of a starburst galaxy, a radio galaxy, and SNR.

\begin{table}[b!]
\caption[Statistics of \gray\ sources at GeV and TeV energies.]
{Statistics of \gray\ sources at GeV and TeV energies.}
\begin{tabular*}{\textwidth}{@{\extracolsep{\fill}}lll}
\sphline
& GeV & TeV \cr
\sphline
Pulsars/Plerions&6 &3 \cr
SNR &0 &2\cr
AGN/Blazars&  60? & 6 \cr
AGN/Radio galaxies &1 &1 \cr
Galaxies & 1 (LMC) &1 (NGC 253)\cr
GRB & 4 & 0\cr
Unidentified& 170 &1 \cr
\sphline
\end{tabular*}
\label{tab:sources}
\end{table}

One of the most unexpected features is the paucity of the overlap
between these two catalogs. While blazars dominate the identified
sources in both energy bands, the brightest TeV blazars are only weak
GeV sources and the brighest GeV blazar is not detected at all at TeV
energies. At GeV energies the pulsed emission from the Crab and Vela
dominates, while at TeV energies only a DC excess from the plerion 
is observed. It is clear that the, as yet, unexplored region between 
10 and 300 GeV will provide the key to linking together GeV and TeV 
\gray\ sky.

With the launch of GLAST we will have the ability to explore at GeV
energies many of the exciting questions raised by the EGRET
observations.  Foremost among these is understanding the nature of the
many unidentified \gray\ sources. The LAT will have the ability to
detect all of these sources with high significance and pinpoint the
locations of these sources with sufficient accuracy that their
identification may be relatively straightfoward. However, we note that
this will not ``solve'' the issue of unidentified \gray\ sources. Thousands 
of new sources will be detected by the LAT. With a few exceptions, such
as pulsars, identifing the nature of a \gray\ source can only be achieved
by observations of these objects over a wide variety of wavebands. The
better localization capabilities of the LAT will make multiwavelength
observations easier (smaller field to survey) and thus more fruitful
than was possible with EGRET sources. With the lowering of the GeV source
sensitivity threshold of the LAT there will be a large increase in both
the numbers and types of sources detected at these energies, thus there
will be a correspondingly increased importance in the role of multiwavelength
observations to identify and study them. 

New GLAST identifications will fall into 3 types: source classes
already known to be \gray\ emitters, sources predicted to be
\gray\ emitters and hopefully some complete surprises. We
anticipate that observations of the GeV sky with the LAT will answer
many of the intriging questions raised by EGRET and we look forward to
discovering the new set of mysteries raised by the next generation
instrument in this waveband.

\acknowledgments

The work by I.~Moskalenko was supported in
part by a NASA Astrophysics Theory Program grant.

\end{document}